\documentclass[conference,anonymous]{IEEEtran}
\IEEEoverridecommandlockouts
\usepackage{cite}
\usepackage{amsmath,amssymb,amsfonts}
\usepackage{algorithmic}
\usepackage{graphicx}
\usepackage{textcomp}
\usepackage{xcolor}

\usepackage[table,xcdraw]{xcolor}

\usepackage{booktabs} 
\usepackage{graphicx} 
\usepackage{array}    
\usepackage{tabularx}  
\usepackage{amsmath}  
\usepackage{color}
\usepackage{soul}
\usepackage{comment}
\usepackage[ruled,vlined]{algorithm2e}
\usepackage[numbers]{natbib}

\usepackage{ragged2e}

\def\BibTeX{{\rm B\kern-.05em{\sc i\kern-.025em b}\kern-.08em
    T\kern-.1667em\lower.7ex\hbox{E}\kern-.125emX}}
\begin{document}

\title{Towards Neurocognitive-Inspired Intelligence: From AI’s Structural Mimicry to Human-Like Functional Cognition
}

\author{\IEEEauthorblockN{Noorbakhsh Amiri Golilarz, Hassan S. Al Khatib, Shahram Rahimi} Department of Computer Science, The University of Alabama, Tuscaloosa, AL\\Email: \{noor.amiri, hsalkhatib, srahimi1\}@ua.edu\\
}
\maketitle

\begin{abstract}
Artificial intelligence has advanced significantly through deep learning, reinforcement learning, and large language and vision models. However, these systems often remain task-specific, struggle to adapt to changing conditions, and cannot generalize in ways similar to human cognition. Additionally, they mainly focus on mimicking brain structures, which often leads to black-box models with limited transparency and adaptability. Inspired by the structure and function of biological cognition, this paper introduces the concept of “Neurocognitive-Inspired Intelligence (NII),” a hybrid approach that combines neuroscience, cognitive science, computer vision, and AI to develop more general, adaptive, and robust intelligent systems capable of rapid learning, learning from less data, and handling minimal complexity based on prior experience. These systems aim to emulate the human brain’s ability to flexibly learn, reason, remember, perceive, and act in real-world settings with minimal supervision. We review the limitations of current AI methods, define core principles of neurocognitive inspired intelligence, and propose a modular, biologically inspired architecture that emphasizes integration, embodiment, and adaptability. We also discuss potential implementation strategies and outline various real-world applications, from robotics to education and healthcare. Importantly, this paper offers a hybrid roadmap for future research, laying the groundwork for building AI systems that more closely resemble human cognition.
\end{abstract}


\section{Introduction}
\label{Introduction}

The pursuit of artificial intelligence that closely mimics human cognition has captivated scientists for decades. From early symbolic systems to the deep learning revolution, AI has achieved unprecedented progress in solving specialized tasks such as image classification, speech recognition, and natural language processing \cite{lecun2015deep} \cite{silver2016mastering} \cite{brown2020language}. However, these advancements are often narrow in scope and brittle when faced with new, ambiguous, or real-world situations. As AI systems are deployed across more complex domains, from healthcare to autonomous systems, the shortcomings of traditional models have become clearer, emphasizing the urgent need for more adaptable, robust, and general-purpose intelligence \cite{marcus2018deep}.

Conventional AI, especially those rooted in deep learning and statistical pattern recognition, heavily rely on large, annotated datasets and are generally optimized for static tasks. While they excel in environments with well-defined inputs and outputs, they often lack key characteristics of human-like intelligence, such as adaptability, context-awareness, causal reasoning, and goal-directed behavior \cite{lake2015human} \cite{geirhos2020shortcut} \cite{bender2020climbing}. These models frequently struggle with out-of-distribution generalization, real-time learning, and long-term temporal reasoning \cite{hassabis2017neuroscience} \cite{zador2019critique}. Furthermore, they often operate as black boxes, lacking transparency and interpretability, which limits their applicability in safety-critical domains \cite{lipton2018mythos}. These limitations highlight the need for a paradigm shift from purely statistical models toward biologically inspired, integrative frameworks that better reflect the dynamic and adaptable nature of human cognition.

In recent years, the concept of brain-inspired intelligence, an AI paradigm inspired by the structural and functional principles of the human brain, has emerged as a promising path toward achieving more general, flexible, and human-aligned intelligence \cite{eliasmith2014use} \cite{goyal2022inductive} \cite{bressler2006operational}. However, brain-inspired methods often focus on architectural mimicry, such as layered neurons, distributed representations, and parallel processing, without fully capturing the functional and dynamic aspects of cognition that support human adaptability and reasoning. Despite these advances, most brain-inspired AI models operate as black boxes, offering limited interpretability, adaptability, and reasoning capabilities. They excel at pattern recognition but struggle with dynamic, uncertain, or novel real-world environments. By drawing on cognitive neuroscience, psychology, and computational modeling, neurocognitive-inspired intelligence aims to replicate core skills such as attention, working memory, perception, mental simulation, and meta-reasoning, often within embodied or interactive contexts \cite{benedek2019toward} \cite {pezzulo2016navigating}. Recent discoveries in brain science, along with advances in neural-symbolic integration, large-scale brain simulations, and attention-based architectures, have brought us closer than ever to building intelligent systems that do not merely learn from static data but understand, reason, and adapt in real time \cite{richards2019deep} \cite{shanahan2010embodiment} \cite{marcus2019rebooting}. These trends suggest that intelligence is not just about optimizing input-output mappings but is an emergent property of dynamic systems that integrate perception, memory, control, and action, precisely the principles embodied by the neurocognitive paradigm.

\begin{figure*}[h]
  \centering
  \includegraphics[width=0.9\linewidth]{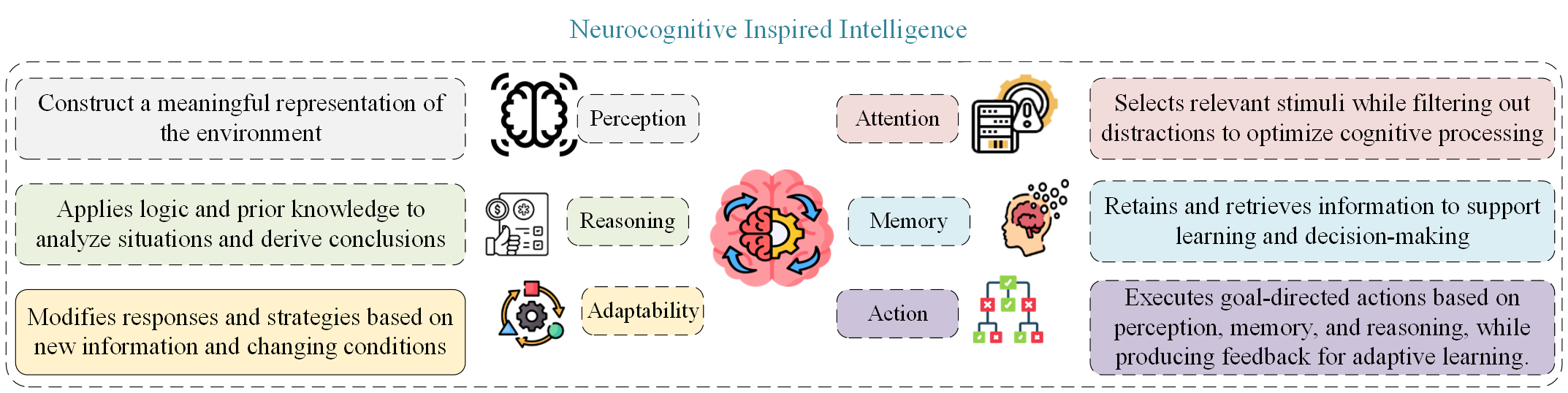}
  \caption{Illustrating neurocognitive inspired intelligence framework, providing a high-level overview of the core components of neurocognitive intelligence and their functional roles. }
  \label{fig:fig_01}
\end{figure*}

In this paper, we propose a comprehensive neurocognitive-inspired intelligence framework that aims to bridge the gap between current AI systems and human-like intelligence. Instead of simply mimicking the brain’s structure, our hybrid conceptual framework draws inspiration from biological cognition, which studies how the brain enables perception, memory, attention, reasoning, and adaptation through interconnected neural processes. These processes include the organization of the brain’s structure and functions, such as attentional modulation, multi-timescale memory systems, neuroplasticity, and predictive coding. Based on these principles, our framework combines key mechanisms from neuroscience and cognitive science into a unified computational architecture. We introduce a multi-level framework that includes a perception-action loop, hierarchical representation learning, attentional filtering, memory-based inference, and self-supervised adaptation (see Fig. \ref{fig:fig_01}). The goal is to develop AI systems capable of real-time reasoning, contextual generalization, and lifelong learning. This proposed framework enhances the theoretical understanding of cognitive AI and paves the way for practical applications in robotics, healthcare, education, and autonomous decision-making systems.
The major contributions of this work are as follows:
\begin{itemize}
    \item We identify and characterize critical limitations of traditional AI from a neurocognitive perspective.
    \item We introduce a biologically inspired, integrative hybrid framework called Neurocognitive-Inspired Intelligence (NII) that models essential cognitive processes such as attention, memory, executive control, and abstraction. In Sections 2 and 3, we refer to this concept as biological cognition, highlighting the fundamental principles from neuroscience and cognitive science upon which the NII framework is based, as formally described in Section 4.
    \item We demonstrate how our proposed framework enables generalization, context-aware reasoning, and adaptive learning in open-world environments.
    \item We include conceptual diagrams and detailed case studies to show the applicability of the neurocognitive framework across various real-world domains, including healthcare, industrial safety, personalized education, and resilient robotics.
\end{itemize}

The rest of this paper is organized as follows. Section II presents a critical review of the limitations in current AI architectures, focusing on issues related to generalization, memory, adaptability, and reasoning. Section III provides the theoretical foundation and interdisciplinary motivations for adopting a biologically grounded cognition approach, including its alignment with brain function and implications for AI. Section IV describes the proposed conceptual neurocognitive-inspired intelligence framework in detail, with architectural diagrams and component explanations. Section V discusses implementation strategies, real-world case studies, and design considerations for practical application. Finally, Section VI summarizes the findings, contributions, and future research directions.

\section{Limitations of Current State-of-the Art}
\label{background}

Although artificial intelligence has made significant progress in recent years, especially with deep learning and large-scale neural networks, these achievements are often limited in scope. While models such as convolutional neural networks (CNNs) and transformers have achieved top performance on benchmarks in vision, language, and speech, these systems still face fundamental limitations that slow progress toward general, adaptable, and human-like intelligence. This section highlights the main limitations of current architectures, grouping them into core cognitive and functional gaps that stand in contrast to human thinking abilities. Table \ref{table:tbl_01} provides a high-level overview comparing the main limitations of traditional AI systems, mapping each limitation to its corresponding biological cognitive capability, the underlying neural and biological foundation, and the practical impact these improvements could bring to real-world applications. Each limitation is then examined in detail in the following subsections.

\begin{table*}[!t]
\caption{Limitations of Conventional AI vs. Biological Cognition, with Biological Inspirations and Real-World Implications}
\label{table:tbl_01}
\centering
\setlength{\tabcolsep}{5pt}
\renewcommand{\arraystretch}{1.15}
\begin{tabular}{|p{3.2cm}|p{3.8cm}|p{3.8cm}|p{3.8cm}|}
\hline
\rowcolor[HTML]{F2CEED}
\textbf{Conventional AI Limitation} &
\textbf{Biological Cognition Capability to Address the Limitations} &
\textbf{Biological Basis} &
\textbf{Impact} \\ \hline
Poor generalization & 
Contextual, semantic, and episodic generalization & 
Hippocampus, association cortex & 
Robust out-of-distribution (OOD) performance \\ \hline
Data inefficiency & 
Few-shot, schema-based learning & 
Hebbian plasticity, schema encoding & 
Learning from limited data \\ \hline
Weak temporal reasoning & 
Working memory, sequential planning & 
Prefrontal cortex, basal ganglia & 
Improved multi-step reasoning \\ \hline
Inflexibility & 
Real-time feedback and goal adaptation & 
Corticothalamic loops, neuromodulation & 
Dynamic, goal-aware AI \\ \hline
Fragmented processing & 
Perception--action--cognition integration & 
Sensorimotor loops, cerebellum & 
Unified cognitive agents \\ \hline
Adversarial fragility & 
Feedback-driven, robust perception & 
Predictive coding, top-down control & 
Safer deployment in critical systems \\ \hline
No metacognition & 
Self-monitoring and uncertainty tracking & 
Prefrontal cortex, default mode network & 
Explainability and introspection \\ \hline
\end{tabular}
\end{table*}


\subsection{Poor Generalization and Out-of-Distribution Robustness}

One of the most recognized limitations of modern AI is its inability to generalize well beyond the training data distribution. While neural networks can interpolate between data points, they often fail to extrapolate to new, unseen scenarios, a behavior that is critical for human-like cognition. For example, models trained on ImageNet might not recognize slightly modified or adversarial versions of familiar objects, exposing brittle perceptual abilities \cite{recht2019imagenet} \cite{geirhos2020shortcut}. This is very different from human vision, which can generalize across different contexts, scales, and even degraded stimuli. Additionally, out-of-distribution (OOD) data remains a major challenge. When encountering new contexts, current models tend to make confident but incorrect predictions. This can have serious consequences for real-world applications such as medical diagnosis or autonomous driving, where safety and accuracy under varying conditions are critical \cite{ovadia2019can}. Human cognition, on the other hand, relies on mechanisms like abstraction, causality inference, and compositional reasoning—skills that current deep learning systems lack. 

\subsection{Data Inefficiency and the Problem of Supervision}

Deep learning systems usually require large, labeled datasets to perform well, which can be costly and impractical in many domains. Humans, on the other hand, can learn effectively from just a few examples by using prior knowledge, context, and reasoning to make inferences. This ability, known as few-shot or one-shot learning, was largely absent from traditional deep learning models \cite{lake2015human}. The reliance on supervised learning also limits AI’s scalability. Labeling data is expensive and sometimes impossible, especially in domains like healthcare or robotics where expert annotations are needed. Approaches like self-supervised and unsupervised learning have emerged as partial solutions, but they still fall short of supervised models in performance and often fail to capture semantic and causal relationships in data \cite{lecun2024path}. Without the ability to form internal representations on its own, as humans do during development, AI remains constrained in its capacity for lifelong learning. 

\subsection{Lack of Temporal Memory and Sequential Reasoning}

Another significant limitation is the lack of explicit long-term memory and temporal reasoning. While transformers have revolutionized sequence modeling with attention mechanisms, they are inherently limited in memory capacity and scalability, especially for very long input sequences. They lack a form of dynamic memory that persists across tasks or sessions which is something that is central to human cognition \cite{dai2019transformer}. Moreover, current models often struggle with tasks that require multi-step reasoning, planning, or understanding of temporal cause-and-effect relationships. For instance, multi-hop question answering or procedural understanding is usually achieved through brute-force memorization rather than structured inference \cite{khot2020qasc}. In contrast, the human brain employs mechanisms like working memory, episodic memory, and executive control to reason about the past, plan for the future, and adapt behavior accordingly. 

\subsection{Inflexibility and Limited Adaptation to New Goals}

Humans excel at rapidly adapting to new goals and constraints, a skill that remains difficult for most AI systems. Most machine learning models are fixed after training: adapting to a new task usually requires fine-tuning or retraining from scratch. In contrast, humans can reconfigure their cognitive processes on the fly, drawing from past experiences and transferring knowledge across different domains \cite{finn2017model}. This inflexibility is especially problematic in dynamic environments, such as robotics or interactive systems, where real-time learning and understanding context are essential. While meta-learning (learning how to learn) and continual learning (learning continuously without forgetting) are promising areas, they still struggle with catastrophic forgetting or issues related to scalability issues \cite{parisi2019continual}. Biological cognition, on the other hand, can excel in open-ended environments because it can encode and revise goals, manage trade-offs, and update beliefs based on incoming stimuli.

\subsection{Fragmented Architectures Lacking Integrated Cognition}

One of the most profound limitations is the absence of integrated cognitive architecture in most AI systems. Most current models tend to specialize in a narrow function, such as perception, language generation, or control, but lack the ability to coordinate multiple cognitive functions in a unified way. In biological systems, cognitive functions such as attention, perception, memory, and reasoning are deeply interconnected and operate as part of a dynamic control system \cite{posner1990attention}. This lack of end-to-end neurocognitive loops means that current AI systems miss critical feedback control, situational awareness, and goal prioritization mechanisms. For instance, perception is not usually modulated by attention in most models, and memory does not dynamically influence reasoning. While cognitive architectures like ACT-R \cite{ritter2019act} or SOAR \cite{laird1987soar} attempt to model such integration symbolically, and systems like AlphaStar \cite{arulkumaran2019alphastar} demonstrate cross-modal functionality, these are still far from biologically grounded intelligence \cite{anderson2009can}\cite{reed2022generalist}.

\subsection{Vulnerability to Adversarial Perturbations}

Contemporary deep learning systems are well-known for their vulnerability to adversarial attacks, where tiny unnoticed changes in input data can lead to incorrect outputs. This vulnerability stems from how deep networks encode features, often in ways that are not resilient to small perturbations. In contrast, humans are remarkably resistant to noisy, distorted, or ambiguous sensory input, thanks to their hierarchical, feedback-driven processing mechanisms \cite{goodfellow2014explaining}.
Efforts to boost robustness through methods like adversarial training, input preprocessing, or certified defenses have yielded progress but have not fully addressed the core fragility. This lack of robustness undermines the deployment of AI in high-stakes areas such as security, finance, and healthcare. A neurocognitive approach could provide greater robustness by mimicking the brain’s ability to filter, contextualize, and reconstruct input amid uncertainty. Table ~\ref{table:tbl_02} compares the human perceptual system with modern AI models, emphasizing key differences in robustness and adaptability. While humans effortlessly manage ambiguous or degraded inputs through context and feedback, most AI systems fail under minimal perturbations. This motivates the pursuit of biologically inspired cognition approaches that replicate the brain’s layered, dynamic, and context-aware processing. 


\newcolumntype{Y}{>{\RaggedRight\arraybackslash}X}

\begin{table*}[!t]
\caption{Comparison of Human vs. AI Response to Adversarial or Noisy Inputs}
\label{table:tbl_02}
\centering
\setlength{\tabcolsep}{5pt}        
\renewcommand{\arraystretch}{1.2}  
\begin{tabularx}{\textwidth}{|Y|Y|Y|Y|}
\hline
\rowcolor[HTML]{F2CEED}
\multicolumn{1}{|c|}{Aspect} &
\multicolumn{1}{c|}{Humans} &
\multicolumn{1}{c|}{Deep Learning Models} &
\multicolumn{1}{c|}{Biological Cognition Implication} \\ \hline
Resilience to Noise &
Robust to visual distortions, occlusions, and background noise &
Often fail with minor, imperceptible perturbations &
Suggests need for hierarchical and feedback-driven processing \\ \hline
Contextual Awareness &
Use contextual cues to disambiguate unclear input &
Lack global understanding; rely on local pixel-level features &
Context integration may improve robustness \\ \hline
Generalization Across Views &
Recognize objects under varied angles, lighting, and occlusion &
Poor generalization outside training distribution &
Embodied learning and multimodal input could aid generalization \\ \hline
Perceptual Stability &
Stable perception over time, despite changes &
High sensitivity to tiny perturbations or data shifts &
Feedback loops and temporal integration can stabilize representations \\ \hline
\end{tabularx}
\end{table*}


\subsection{Absence of Common Sense and World Models}

Current AI systems lack common sense reasoning, which is the intuitive understanding of physical and social realities that humans develop early in life. For example, large language models can produce grammatically correct but logically nonsensical text. They often struggle to reason about object permanence, causality, or emotional nuance \cite{marcus2002next}. Building world models \cite{ha2018world}, internal simulations of how the world functions, remains an open challenge in AI. Humans use such models to imagine outcomes, predict others’ behavior, and plan future actions. Although efforts like model-based reinforcement learning or neural-symbolic integration aim to close this gap, a scalable, adaptable world modeling approach remains elusive \cite{ha2018world}. The limitations outlined above reveal fundamental flaws in traditional AI architectures, especially their inability to generalize, adapt, and reason in a human-like manner. These gaps highlight the growing need for a new class of intelligent systems, inspired by the core capabilities of biological cognition.

\section{The Need for Neurocognitive-Inspired Intelligence}
\label{nuero_int}

To overcome the fundamental limitations of current artificial intelligence systems, it is essential to look beyond engineering-inspired architectures and draw inspiration from the most versatile intelligent system we know: the human brain. This section details the scientific reasoning, core mechanisms, and design principles behind biologically grounded cognition approaches. Although we refer to the overall conceptual framework as Neurocognitive Intelligence (NI) in Section 4, the concepts introduced here establish the foundation for its design, explaining why each component is important and which biological principles influenced them.
\subsection{Biological Plausibility and Embodied Cognition}
The human brain has evolved over millions of years to function as a strong, adaptable, and general-purpose information-processing system. It combines sensory input, encodes memories, learns from limited feedback, and enables complex planning and social interactions. Unlike modern AI models that depend on fixed architectures trained for specific tasks, the brain constantly reorganizes its internal representations based on context and experience \cite{hassabis2017neuroscience} \cite{eliasmith2014use}. Biologically rooted cognition models might imitate this flexible and dynamic nature through architectures inspired by the brain's functional organization, such as hierarchical processing, lateral inhibition, predictive coding, and recurrent loops.
Cognition is not only an isolated computational process but is also deeply rooted in sensory and motor interactions with the physical world. Theories of embodied cognition suggest that intelligent behavior emerges from the ongoing interaction between an agent's body, brain, and environment \cite{clark1999embodied}. Biologically grounded cognition models incorporate this idea by using sensorimotor feedback to connect abstract concepts to real-world experiences, thus improving generalization and adaptability. For example, robots equipped with neuro-inspired learning modules can learn about affordances and object functions through physical manipulation instead of relying on symbolic rules \cite{cangelosi2015developmental}.

\subsection{Perception-Action Loops and Predictive Processing}
One of the most significant shortcomings of current AI is the absence of a closed-loop system that connects perception to action through reasoning and memory. In biological systems, perception is active rather than passive, guided by goals, expectations, and past experience. The perception-action loop enables organisms to adaptively decide what to focus on, which actions to perform, and how to learn from the consequences \cite{pezzulo2016navigating}. Biologically grounded cognition models aim to imitate this by linking sensory processing modules with decision-making and motor control modules in a continuous feedback loop, supporting real-time learning and context-aware behavior.
The human brain is thought to function as a hierarchical prediction system that constantly makes hypotheses about incoming sensory input and minimizes prediction errors through feedback \cite{friston2010free}. This predictive coding model enables efficient data processing and reliable inference in noisy environments. Biologically grounded cognition models can incorporate this by using generative models and recurrent feedback to forecast outcomes and adapt internal models dynamically. For example, variational autoencoders and generative adversarial networks provide computational tools for applying such predictive frameworks in AI \cite{kingma2013auto}.

\subsection{Memory in Human and Artificial Intelligence}
Unlike current AI systems that often rely on short-term context windows (e.g., transformers), the human brain stores both episodic (event-based) and semantic (fact-based) memories that can be retrieved based on context. The hippocampus is crucial for consolidating episodic memories, while the neocortex manages semantic abstraction \cite{squire1991medial}. Biological cognition mimics this by using external memory modules or neuromorphic memory circuits \cite{pershin2011neuromorphic} that enable long-term storage and retrieval of task-relevant data. These mechanisms are foundational for creating AI agents capable of lifelong learning without experiencing catastrophic forgetting \cite{kirkpatrick2017overcoming}.
Human intelligence is characterized by the ability to learn how to learn, known as meta-learning. Through prior experience, humans develop strategies to acquire new skills quickly and efficiently. Biologically grounded cognition models incorporate meta-learning mechanisms to enable rapid adaptation in new situations. This is often achieved using meta-gradient techniques, dynamic neural reconfiguration, or modular learning pathways that mimic the flexibility of the prefrontal cortex \cite{wang2016learning} \cite{miconi2018differentiable}. Such systems show promise in few-shot and zero-shot learning tasks.

\subsection{Attention in Cognitive Systems}
Attention is the brain’s mechanism for filtering relevant information from overwhelming sensory input. Biological systems use both bottom-up (stimulus-driven) and top-down (goal-directed) attention strategies to prioritize stimuli \cite{itti2002model}. While transformer models have made progress in simulating attention through self-attention, most lack the dynamic, context-sensitive control and biologically grounded multimodal integration found in human cognition. The neurocognitive intelligence framework aims to lay the foundation for developing adaptive attention mechanisms that modulate processing based on task goals, emotional salience, and contextual cues. Table \ref{table:attention_ai_vs_bio} provides a comparative overview of how state-of-the-art multimodal transformers (e.g., CLIP \cite{radford2021learning}, BLIP \cite{li2022blip}, Flamingo \cite{alayrac2022flamingo}) differ from biologically grounded cognition models in terms of neural grounding of attention mechanisms, architecture, contextual modulation, and adaptive control mechanisms.

Another key feature of natural intelligence is the seamless integration of information across various sensory modalities. For instance, vision and hearing often work together to enhance perception and resolve ambiguities in input. Neurocognitive models utilize multimodal neural structures, such as cross-modal transformers \cite{yan2023cross} and sensor fusion layers \cite{elmenreich2002introduction} \cite{yeong2021sensor} to imitate this ability, leading to richer, more grounded understanding and more precise decision-making in complex environments \cite{baltruvsaitis2018multimodal}. Table \ref{table:multimodal_integration_bio} demonstrates how different modality combinations support specific cognitive functions and their related applications in neurocognitive intelligence, highlighting the importance of sensor fusion for context-aware and adaptive systems.

\subsection{Goal-Directed Reasoning and Executive Control}
While many deep learning systems are reactive and feedforward in nature (taking input, computing an output, and moving on without revisiting or refining their decisions, planning, or considering long-term consequences), humans plan future actions based on current goals, anticipated outcomes, and prior experience. A biologically grounded cognition model can implement this through symbolic planning modules \cite{konidaris2018skills}, reinforcement learning controllers \cite{schmidhuber2015learning}, and hierarchical policy networks \cite{huang2020joint}. These mechanisms work together to support goal-oriented behavior, enabling AI agents to perform multistep tasks with intermediate decision points \cite{silver2016mastering}. Executive functions such as task-switching, inhibition, and conflict monitoring are essential for intelligent behavior. The brain’s prefrontal cortex handles these functions, allowing flexible responses in the face of uncertainty or distraction \cite{miller2001integrative}. These models can simulate such control by incorporating neural gating mechanisms \cite{gu2020improving} and context-sensitive modulators \cite{sarkar2001small} that regulate information flow and suppress irrelevant responses. These improvements make AI systems more robust, safe, and easier to interpret. 

\subsection{Neuroplasticity and Real-Time Adaptation}
Real-time adaptation is a vital gap in current AI. Humans exhibit neuroplasticity, which enables them to reorganize neural connections in response to new information, experiences, learning, or injury. Inspired by this, biologically grounded cognition approaches are exploring continuous learning models and local synaptic updates that allow AI to adjust its internal states without needing full retraining \cite{zenke2017continual}. These systems employ mechanisms such as Hebbian learning \cite{gerstner2002mathematical}, synaptic consolidation \cite{clopath2012synaptic} \cite{bramham2005bdnf}, and attention-based adaptation \cite{zuo2021attention} to emulate the flexibility of biological intelligence. The limitations of current AI paradigms, ranging from poor generalization and lack of memory to brittle perception, and limited reasoning, highlight the need for a new approach to intelligent systems. Neurocognitive intelligence (detailed in Section 4) offers a promising path forward by integrating principles from biological and cognitive sciences into artificial systems. By combining perception-action loops, predictive processing, long-term memory, attention control, and goal-oriented planning, neurocognitive architecture brings us closer to machines that can think, learn, and adapt like humans.

\newcolumntype{Y}{>{\RaggedRight\arraybackslash}X}

\begin{table*}[!t]
\caption{Comparison of Attention Mechanisms in Multimodal AI vs.\ Biological Cognition}
\label{table:attention_ai_vs_bio}
\centering
\setlength{\tabcolsep}{6pt}
\renewcommand{\arraystretch}{1.15}
\begin{tabular}{|p{3.8cm}|p{5.8cm}|p{5.8cm}|}
\hline
\rowcolor[HTML]{F2CEED}
\textbf{Cognitive Dimension} &
\textbf{CLIP / BLIP / Flamingo} &
\textbf{Biological Cognition} \\ \hline
Multimodal Fusion &
Performs basic image--text fusion using self- or cross-attention &
Integrates vision, language, and tactile input via embodied, adaptive fusion \\ \hline
Task-Driven Attention &
Attention is static post-pretraining &
Dynamically modulated by task goals, memory, and feedback \\ \hline
Context-Sensitive Salience &
No top-down influence from emotional or situational cues &
Attention guided by memory, goals, and environment \\ \hline
Iterative Feedback Loops &
Largely feedforward; limited or no recurrent processing; not involving iterative loops between vision, language, memory, and reasoning &
Incorporates closed-loop feedback between modules (e.g., memory $\leftrightarrow$ reasoning $\leftrightarrow$ attention) \\ \hline
Biological Grounding &
Mathematical abstraction of attention mechanisms &
Inspired by prefrontal--parietal attentional circuits in the brain \\ \hline
Embodied Interaction &
No sensorimotor awareness; cannot shift attention based on bodily or environmental interactions &
Attention modulated by physical interaction, proprioception, and tactile cues \\ \hline
Adaptation to Novelty or Failure &
Lacks mechanisms for adjusting attention in real time &
Attention shifts dynamically to anomalies, uncertainty, or novelty \\ \hline
\end{tabular}
\end{table*}

\begin{table*}[!t]
\caption{Multimodal Integration in Biological Cognition}
\label{table:multimodal_integration_bio}
\centering
\setlength{\tabcolsep}{6pt}
\renewcommand{\arraystretch}{1.15}
\begin{tabularx}{\textwidth}{|Y|Y|Y|}
\hline
\rowcolor[HTML]{F2CEED}
\multicolumn{1}{|c|}{Modalities Combined} &
\multicolumn{1}{c|}{Function Enabled} &
\multicolumn{1}{c|}{Example Application} \\ \hline

Vision + Tactile &
Physical interaction understanding &
Resilient, multitask robotic manipulation \\ \hline

Vision + Auditory &
Context disambiguation, event prediction &
Human--robot communication, surveillance \\ \hline

Vision + Proprioception &
Spatial awareness, motor planning &
Navigation in dynamic environments \\ \hline

Vision + Language &
Semantic grounding, contextual reasoning &
Multimodal assistants, VQA systems \\ \hline

Auditory + Language &
Speech comprehension, sentiment detection &
Conversational agents, assistive tech \\ \hline

Vision + Thermal + Tactile &
Safety and anomaly detection &
Industrial inspection, healthcare robots \\ \hline
\end{tabularx}
\end{table*}

\section{Overview of the Neurocognitive-Inspired Intelligence (NII) Architecture}
\label{concept}

Designing artificial intelligence systems that mimic human cognition requires a shift from narrowly trained models to architectures inspired by how the brain integrates perception, memory, attention, reasoning, and action. This section proposes a hybrid neurocognitive intelligence framework that connects these functions, focusing on structural mimicry of biological systems and creating functional counterparts to dynamic cognitive processes. Neurocognitive Inspired Intelligence (NII) is a paradigm that blends ideas from neuroscience, cognitive psychology, computer vision, and artificial intelligence to create models that perceive and classify reason, remember, adapt, and act in dynamic environments. Such models aim to overcome the limitations previously outlined, including issues with generalization, context awareness, lifelong learning, and real-time adaptability.

\subsection{Overview of the Neurocognitive-Inspired Intelligence (NII) Architecture}

At the core of the neurocognitive framework is a looped cognitive cycle that closely resembles processes observed in human cognition. This framework models brain-inspired cognitive processing through a hierarchical structure that enables intelligent agents to perceive, reason, adapt, and act in changing environments. The architecture consists of seven interconnected and biologically inspired modules: Perception Unit, Attention Mechanism, Memory Module, Learning Module, Reasoning Engine, Adaptation Layer, and Action and Decision Execution Unit. These components work together through recurrent, top-down, and bottom-up pathways that mimic essential neurocognitive functions. 

\textbf{Input modalities} include heterogeneous, multimodal data such as visual streams, auditory cues, tactile information, proprioception, and symbolic inputs (e.g., text or goals). These are processed by the \textbf{Perception Unit}, which converts raw signals into structured, mid-level representations using hierarchical feature encoding and abstraction. The perception unit not only detects and encodes environmental signals but also engages in active sensing, incorporating top-down feedback to guide where, when, and how to gather additional information—similar to saccadic vision or attention shifts in biological agents. Once perceptual information is generated, it is routed through the \textbf{Attention Mechanism}, which filters and prioritizes the most salient or goal-relevant inputs. This module dynamically allocates cognitive resources by utilizing both bottom-up salience detection and top-down goal modulation. It ensures that only contextually relevant signals are passed forward to memory and reasoning systems, thereby reducing noise and enhancing cognitive efficiency. Attention also acts as a routing interface, determining which sensory channels or memory resources should be active for a specific task.

The \textbf{Memory Module} stores and retrieves both short-term working memory and long-term episodic or semantic knowledge. These memory traces are essential for maintaining continuity over time, drawing from past experiences, and providing grounding for inference and planning. The memory system supports associative retrieval, schema abstraction, and simulated memory consolidation, enabling the agent to generalize across tasks and contexts while avoiding catastrophic forgetting. The \textbf{Learning Module} updates stored representations and task heuristics based on feedback and new experience, writing refined abstractions back to memory and providing improved priors to downstream inference. Building on perceptual inputs, memory recall, and learning outputs, the \textbf{Reasoning Engine} performs both subsymbolic inference (e.g., via embeddings and neural activations) and symbolic reasoning (e.g., rule-based logic, planning graphs, or causal modeling). It synthesizes contextual information, draws conclusions, evaluates alternatives, and creates action plans based on current goals. This dual-layered reasoning process allows the system to respond efficiently in routine situations while engaging in deliberative planning in complex or new scenarios. The \textbf{Adaptation Layer} plays a key role in adjusting the cognitive architecture based on internal confidence, external feedback, and changing environmental conditions. It functions as a meta-controller that monitors performance, prediction errors, and system-level uncertainty, \textbf{routing evaluative signals to the Learning Module and} initiating updates across modules as needed. Through mechanisms such as meta-learning, continual learning, and strategic reconfiguration, this layer updates learning rates, reconfigures cognitive pathways, and initiates memory consolidation or attention shifts when needed. It enables the system to exhibit lifelong learning, recover robustly from failures, and scale generalization across different domains.

\begin{figure}
  \centering
  \includegraphics[width=0.7\linewidth]{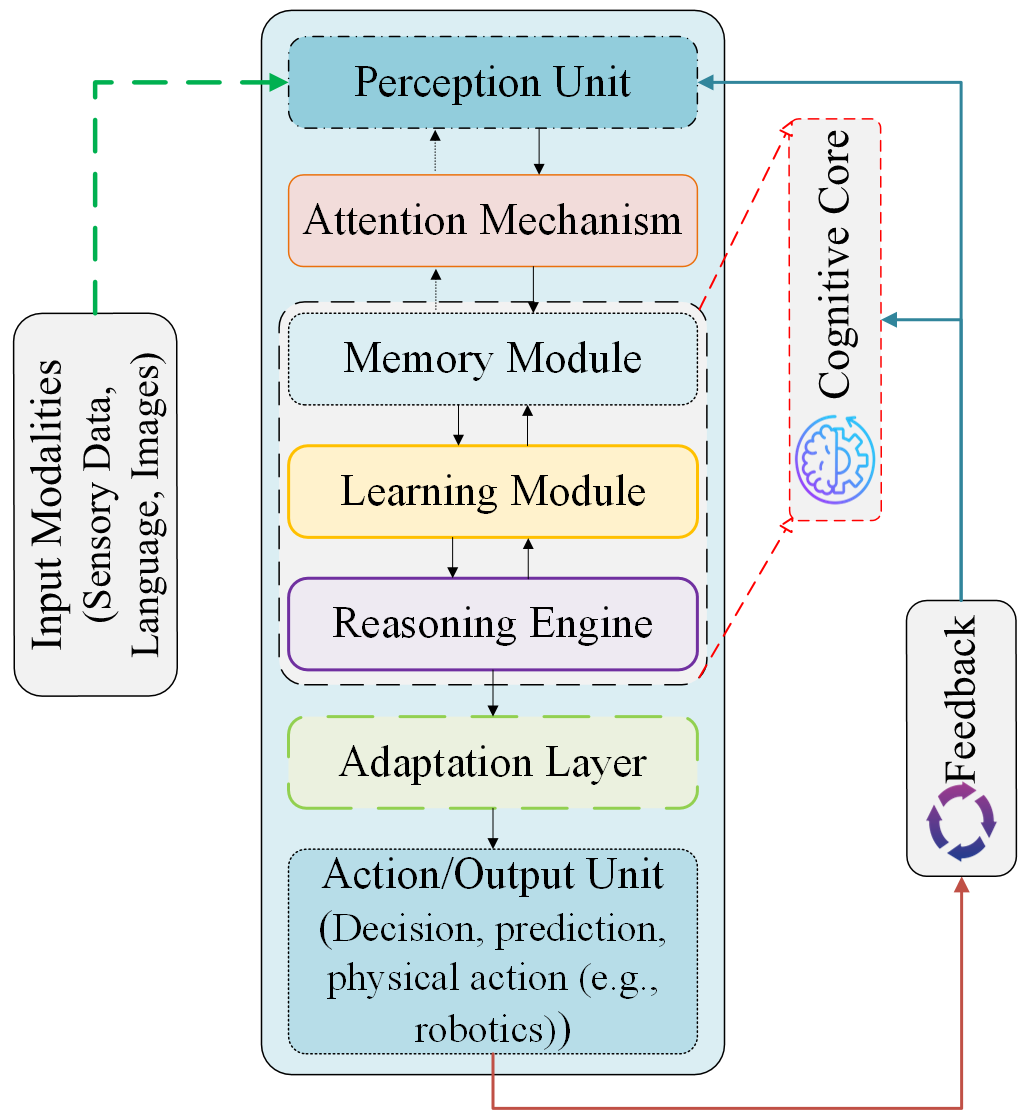}
  \caption{Neurocognitive-Inspired Intelligence (NII) framework with a dedicated \textbf{Learning Module}. Inputs (e.g., sensory data, language, images) are encoded by the \textbf{Perception Unit} and prioritized by the \textbf{Attention Mechanism}. Salient information is stored/retrieved by the \textbf{Memory Module} and refined by the \textbf{Learning Module}, which updates representations and priors used by the \textbf{Reasoning Engine}. The \textbf{Adaptation Layer} modulates strategies based on prediction error, confidence, and context, and the \textbf{Action/Output Unit} executes decisions. Feedback from actions is routed back to perception and into the cognitive core (memory--learning--reasoning), enabling continual improvement, error correction, and context-aware control.}
  \label{fig:nii}
\end{figure}

The \textbf{Action and Decision Execution Unit} transforms internal cognitive output into behavior, physical movement in a robotic system, verbal responses in a conversational agent, or policy choices in a decision-making system. This unit selects goal-driven actions, adjusts execution based on confidence levels, and assesses the results of its outputs. The outcomes of decisions feed back into perception, closing the cognitive loop and supporting ongoing refinement of internal models. These six modules together form a recursive, hierarchical, embodied intelligence system that enables situated cognition, real-time adaptation, contextual reasoning, and interactive decision-making. Each module is inspired by biological processes and based on computational design principles that allow flexible integration, modular reuse, and explainable behavior. Fig. \ref{fig:nii} shows the full neurocognitive architecture, illustrating the directional flow, interactions between modules, and feedback-driven control that define this biologically grounded approach to general intelligence. The framework combines distinct yet interconnected cognitive modules, including perception, attention, memory, reasoning, adaptation, and action, each modeled after corresponding functions in the human brain. It demonstrates how perceptual input passes through mid-level processing and salience detection, ultimately guiding memory and reasoning pathways, and influencing decision-making and action execution. The architecture also emphasizes continuous feedback loops, adaptive modulation from the adaptation layer, and context-aware control systems that support real-time learning and goal alignment. Each of these components is detailed in the upcoming subsections. Together, these components form a unified, dynamic cognitive loop designed to support flexible, embodied, and human-like AI behavior. 

Figure~\ref{fig:nii} depicts the full neurocognitive architecture, highlighting the directional flow, inter-module interactions, and feedback-driven control that characterize this biologically grounded approach to general intelligence. The framework integrates distinct yet interconnected cognitive modules, including perception, attention, memory, learning, reasoning, adaptation, and action, each inspired by corresponding functions in the human brain. It illustrates how perceptual input flows through mid-level processing and salience detection, into memory and reasoning pathways, ultimately informing decision-making and action execution. Architecture also emphasizes continuous feedback loops, adaptive modulation from the adaptation layer, and context-sensitive control mechanisms that enable real-time learning and goal alignment. Each of these components is described in detail in the future subsections. Together, these components form a unified, dynamic cognitive loop, designed to support flexible, embodied, and human-like AI behavior.

Rather than statically processing data end-to-end, the architecture dynamically reuses cognitive subsystems within a feedback loop. For example, memory retrieval can influence reasoning paths, and attention can shift based on earlier inference results, mirroring the top-down modulation common in human cognition \cite{summerfield2014expectation}. Such architecture forms a closed-loop system that allows for responsiveness, error correction, and continuous adaptation. The neurocognitive cycle also supports multitasking and switching between different contexts. It doesn't depend on task-specific optimization but uses reusable, domain-general modules that can be combined flexibly. This adaptability is important for solving tasks with uncertainty or changing constraints, a capability that remains difficult for most traditional AI systems.

\subsection{Perception Unit}

\begin{figure}
  \centering
  \includegraphics[width=1.0\linewidth]{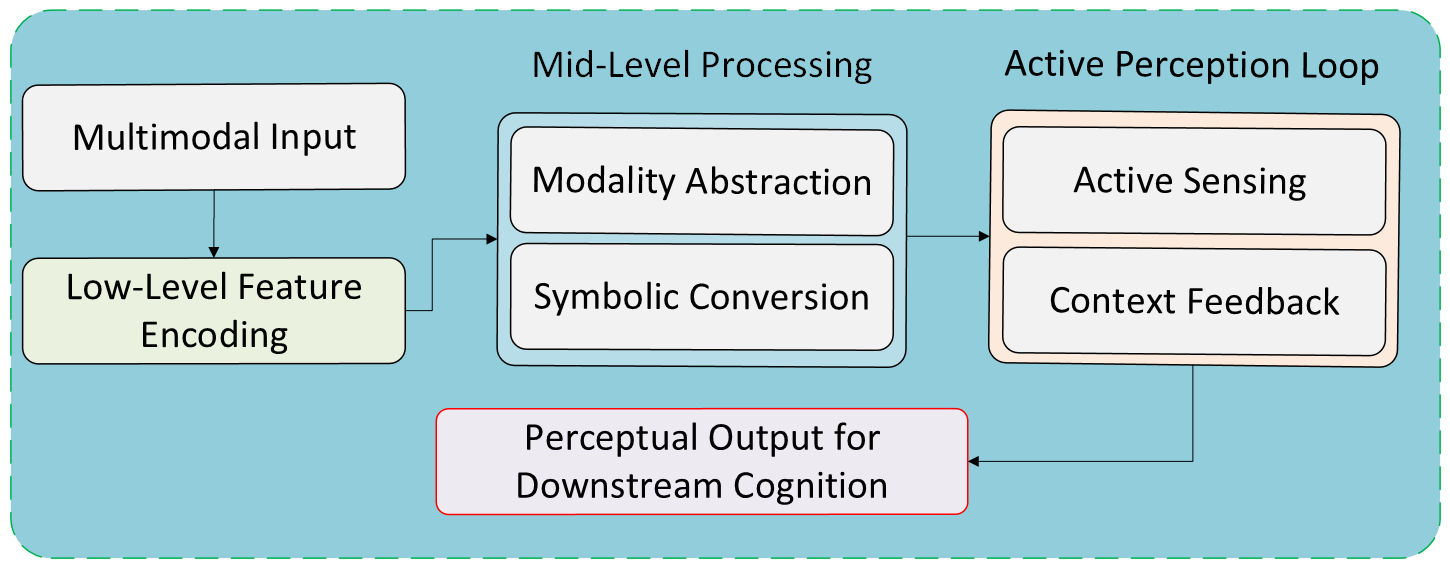}
  \caption{Perception unit. A hierarchical module that transforms raw sensory inputs (visual, auditory, tactile, proprioceptive) into mid- to high-level abstractions via biologically inspired encoders, active perception, and symbolic grounding.}
  \label{fig:fig_03}
\end{figure}

The Perception Unit (as shown in Fig. \ref{fig:fig_03}) within the neurocognitive intelligence architecture serves as the entry point for environmental and sensory data, converting raw multimodal inputs into organized internal representations. Unlike traditional AI systems that rely on static, feedforward pipelines for data intake, this unit is inspired by biological processes and features a hierarchy that performs layered abstraction and symbolic integration. This design allows for flexible, context-aware interpretation of stimuli. The architecture replicates the functions of sensory cortices and the posterior parietal cortex, which process and combine signals from visual, auditory, tactile, and proprioceptive channels.

At the lowest level, sensory input, such as pixels from a camera, audio signals from microphones, or pressure maps from tactile sensors, is encoded by specialized low-level processors. These may include convolutional filters, wavelet transformations, or neuromorphic encoders, each designed to extract basic features like edges, textures, and frequency bands. Instead of relying on a single neural backbone, the perception unit supports modular input encoders optimized for each modality, enabling shared processing when helpful (e.g., early vision-tactile fusion) and separate pipelines when needed. Above the feature extraction layer, the unit builds increasingly abstract representations, from object detection and semantic segmentation to scene graphs and symbolic representations of relationships. This mirrors the visual cortex's progression from simple cells in V1 detecting edges \cite{elder2004psychophysical} to IT cortex representations of full objects \cite{cadieu2014deep}. Symbol grounding mechanisms convert perceptual features into structured concepts, allowing objects like 'cup' or 'metallic texture' to be reasoned about symbolically by higher-level units.

An important innovation in this unit is active perception. Inspired by human visual search and tactile exploration, the system can direct its attention or sensors to acquire new information iteratively. For example, in occluded scenes, eye movement or tactile scanning may be triggered to resolve ambiguity. This closed-loop interaction supports perception-as-inference, a probabilistic reasoning process that combines prior knowledge and feedback to continually refine its internal state. The perceptual module connects directly with memory and attention units to prioritize relevant input channels and recall past context when interpreting new stimuli. 

\subsection{Attention Mechanism: Dynamic Modulation and Contextual Salience}

The attention mechanism in a neurocognitive framework acts as a dynamic gatekeeper of cognition, controlling what information enters working memory, how resources are allocated, and which features influence reasoning and action. Unlike standard attention in transformer-based deep learning, which statically computes relevance via learned weight matrices, the neurocognitive attention module is task-sensitive, feedback-driven, and rooted in biological cortical control mechanisms. In biological systems, attention results from an interaction between top-down control signals, driven by goals, expectations, and memories, and bottom-up cues such as novelty, movement, or emotional salience. These processes are managed by the prefrontal cortex, posterior parietal cortex, and thalamic relay centers, enabling the brain to flexibly select and shift attention in response to changing internal states and external stimuli.

\begin{figure}
  \centering
  \includegraphics[width=0.53\linewidth]{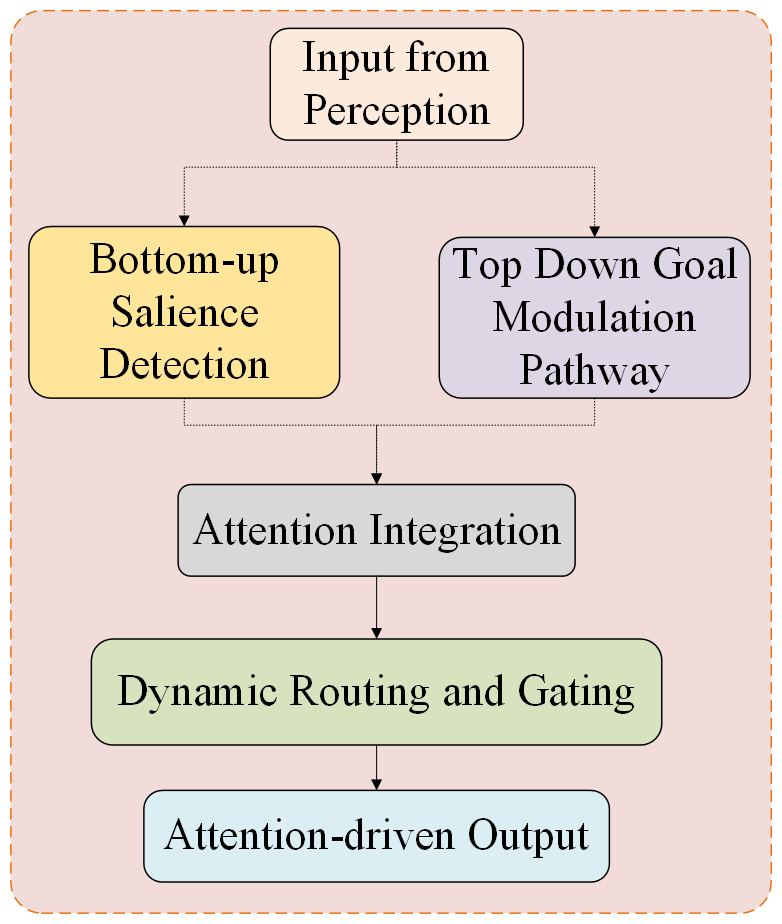}
  \caption{Attention mechanism. A bidirectional, context-sensitive attention controller, highlighting how top-down goals and bottom-up salience modulate cognitive resource allocation.}
  \label{fig:fig_04}
\end{figure}

The neurocognitive attention module mimics this dual structure. First, top-down modulation is key: task demands, past reasoning results, or retrieved memory traces influence what features to focus on at any given time. For instance, during a navigation task, attention may be directed to visual landmarks if spatial disorientation is detected. This goal-oriented modulation adds cognitive flexibility that fixed-attention mechanisms in most transformer models lack. Second, bottom-up salience detection captures stimulus-driven attention shifts. Salient cues, like sudden movement in a scene or a new linguistic token, trigger attentional shifts even without predefined goals. This aligns with the brain’s ability to reprioritize processing based on unexpected environmental signals, enhancing reactivity and robustness. Importantly, this attention is not limited to spatial domains (e.g., regions of an image) but also extends to temporal aspects (e.g., which past event to recall), semantic relevance (e.g., which concept matters most), and procedural steps (e.g., which cognitive subroutine to activate). For example, in decision-making under uncertainty, the framework might shift attention backward in time to reassess previously stored observations or forward toward possible outcomes by simulating scenarios.

Architecturally, such a framework requires components beyond traditional attention heads. It might use adaptive routing networks to dynamically reconfigure computational graphs based on salience and goal priors. Neurobiologically inspired mechanisms such as spiking attention maps or neuromodulatory gating functions (similar to dopamine or acetylcholine signaling) \cite{brzosko2019neuromodulation} \cite{zenke2018superspike} can be integrated to adjust gain and noise thresholds across different modalities and levels of abstraction. The attention mechanism also works bidirectionally: it not only filters perceptual input but also modulates reasoning pathways, memory encoding priorities, and even motor planning. For example, when reasoning about a conflicting outcome, the system may shift attention to counterfactual cues stored in episodic memory, aiding deliberation and reappraisal. In the NI framework, attention is not a static relevance matrix but a flexible, hierarchical controller that enables real-time prioritization across sensory, cognitive, and motor domains. It acts as the link between task demands, environmental cues, and internal memory systems, crucial for scalable, context-aware cognition in intelligent agents. Figure \ref{fig:fig_04} illustrates the dual structure of the attention mechanism. 

\subsection{Memory Module: Dynamic Representation, Retention, and Retrieval}

Memory serves as a fundamental component of neurocognitive intelligence, serving as both a storage of experiences and a dynamic processor that aids decision-making, reasoning, and adaptation. Unlike traditional AI systems, where memory is often implicit and distributed across parameters, our framework's proposed memory module is an explicit, structured system that includes both short-term (working) and long-term (episodic and semantic) memory. This dual-system approach is inspired by neurobiological evidence showing that separate brain regions, such as the prefrontal cortex for working memory and the hippocampus for episodic encoding, manage different types of memory with unique temporal and functional traits.

\begin{figure}
  \centering
  \includegraphics[width=0.75\linewidth]{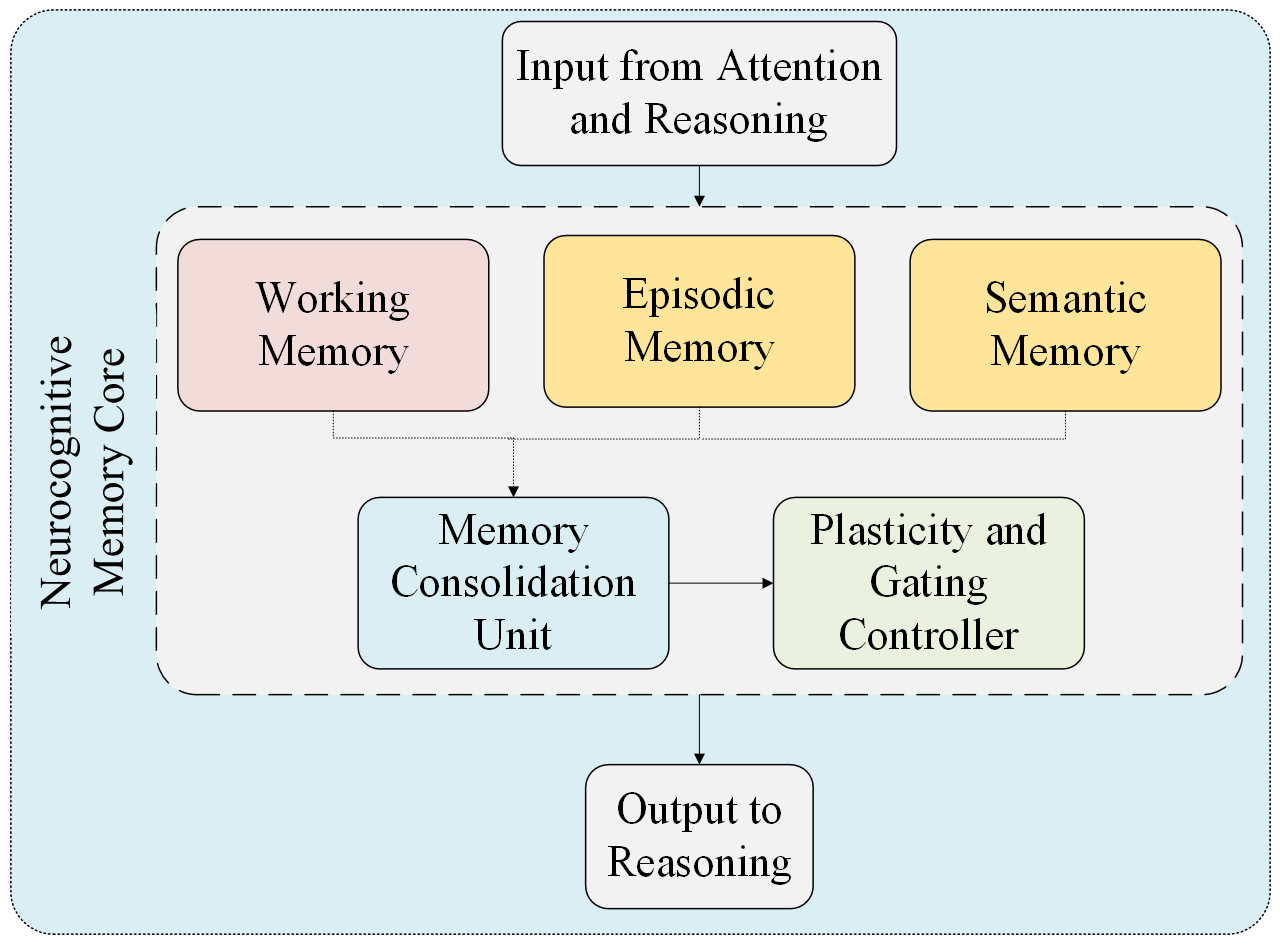}
  \caption{Memory module. A dual-memory system inspired by biological memory structures, supporting working memory, episodic recall, semantic abstraction, and lifelong learning through consolidation and synaptic plasticity.}
  \label{fig:fig_05}
\end{figure}

Working memory \cite{baddeley2020working} in this context acts as a temporary buffer that retains contextually relevant information over short periods to support complex reasoning, task execution, and attention control. It supports the active maintenance and manipulation of representations needed for goal-oriented actions, such as updating a plan with new perceptual input or tracking intermediate states in a multi-step task. This subsystem is designed as a volatile but highly accessible memory bank capable of quick read and write operations guided by attentional control signals. Inspired by neurophysiological and neurocognitive research, we model this process using neural gating mechanisms that control what enters and exits working memory based on relevance, salience, or goal alignment. Long-term memory, on the other hand, stores consolidated experiences and generalized abstractions that enable an agent to learn from the past and apply that knowledge to new situations. It consists of two main subcomponents: episodic memory \cite{tulving2002episodic}, which encodes temporally ordered, context-rich representations of past events; and semantic memory \cite{kumar2021semantic}, which captures structured knowledge and concepts that go beyond individual experiences. Together, these subcomponents allow the agent to recall previous outcomes, identify causal relationships, and use learned schemas to solve unfamiliar problems. Memory retrieval is associative and sensitive to context, often triggered by perceptual or attentional cues that match current input with stored traces. This process enables analogical reasoning and one-shot learning, connecting new stimuli to known solutions.

Crucially, our framework incorporates mechanisms for memory consolidation \cite{squire2015memory} and reorganization, paralleling the biological processes of memory stabilization observed during sleep or rest. In practice, this involves the selective transfer of representations from working memory to long-term storage, potentially utilizing offline simulation or replay mechanisms similar to hippocampal reactivation. These mechanisms enable the system to refine its knowledge base without requiring continual online exposure, thereby supporting both data efficiency and stability. To avoid catastrophic forgetting (a pervasive issue in continual learning), the memory module integrates synaptic consolidation principles and neuromodulatory signals that prioritize the retention of task-critical knowledge. This is implemented through plasticity-aware learning rules \cite{abraham1996metaplasticity}, such as Hebbian updates with adaptive regularization, and memory isolation mechanisms that compartmentalize representations across tasks. The result is a memory system capable of both retaining core competencies and acquiring new skills flexibly. Furthermore, the memory module interacts bidirectionally with the attention and reasoning systems. Memory not only stores the outcomes of inference but also primes attention by modulating the salience of incoming stimuli based on prior experiences. It also influences reasoning by providing episodic context or factual knowledge to guide the inference engine’s symbolic or neural computations. This closed-loop interaction allows the agent to re-evaluate decisions, reflect on past outcomes, and adjust its internal models based on experience.

In sum, the memory module in the neurocognitive intelligence framework goes beyond passive storage. It represents a biologically inspired, dynamic foundation for learning, reasoning, and adaptation. By mimicking the structural organization, encoding mechanisms, and control processes of human memory systems, this module allows AI agents to operate in temporally extended, uncertain, and semantically rich environments. It is a fundamental element for achieving true lifelong learning and contextual generalization, abilities that are still difficult to attain with current machine intelligence. As shown in Fig. \ref{fig:fig_05}, the memory module combines short-term and long-term memory components, supporting associative recall, schema abstraction, and memory consolidation.

\subsection{Learning Module: Continual Knowledge Acquisition and Experience Integration}

The Learning Module in our framework connects the Memory Module and the Reasoning Engine. While memory stores past experiences and reasoning uses that knowledge for inference and decision-making, the learning module controls how these stored representations change over time. This separation reflects the biological difference between memory retrieval and synaptic plasticity–driven learning found in hippocampal-cortical interactions. Functionally, the Learning Module has three complementary roles. First, it identifies patterns and abstractions from episodic traces stored in memory, turning raw experiences into generalized knowledge that supports reasoning across tasks and domains. Second, it incorporates feedback and prediction-error signals from the Adaptation Layer and Action Unit to update stored representations, improving causal links, schemas, and rules as the agent interacts with the environment. Third, it adjusts its reasoning strategies, such as changing how evidence is weighted or which inference pathways are favored, allowing the system to enhance its decision-making abilities over time.

The module interacts bidirectionally with both Memory and Reasoning. It queries memory to retrieve relevant experiences for comparison and writes back newly learned abstractions, supporting consolidation and preventing catastrophic forgetting through selective rehearsal and plasticity-aware updates. It also provides the Reasoning Engine with more refined priors, causal links, or heuristics, while using reasoning outcomes, such as inferred explanations or simulated counterfactuals, as additional learning signals. This module is inspired by biological processes like hippocampal replay, dopaminergic reward-prediction error coding, and prefrontal-guided skill acquisition. In computational terms, it can be implemented as a hybrid continual-learning system that combines experience replay, synaptic consolidation, meta-learning, and reinforcement learning with structured memory augmentation. By explicitly representing the learning process as a separate module, the framework gains a dedicated mechanism for lifelong knowledge acquisition and reasoning strategy refinement, a capacity that is largely missing in traditional end-to-end AI systems.

\subsection{Reasoning and Inference Engine}

The Reasoning and Inference Engine is the cognitive core responsible for interpreting stored knowledge, integrating it with current perceptual inputs, and generating goal-directed conclusions, decisions, or predictions. Within the core, it operates \emph{downstream of the Memory–Learning complex}: Memory provides episodic and semantic content, the Learning Module maintains and refines those representations and heuristics over time, and the Reasoning Engine consumes and, in turn, informs both.

\begin{figure}
  \centering
  \includegraphics[width=0.8\linewidth]{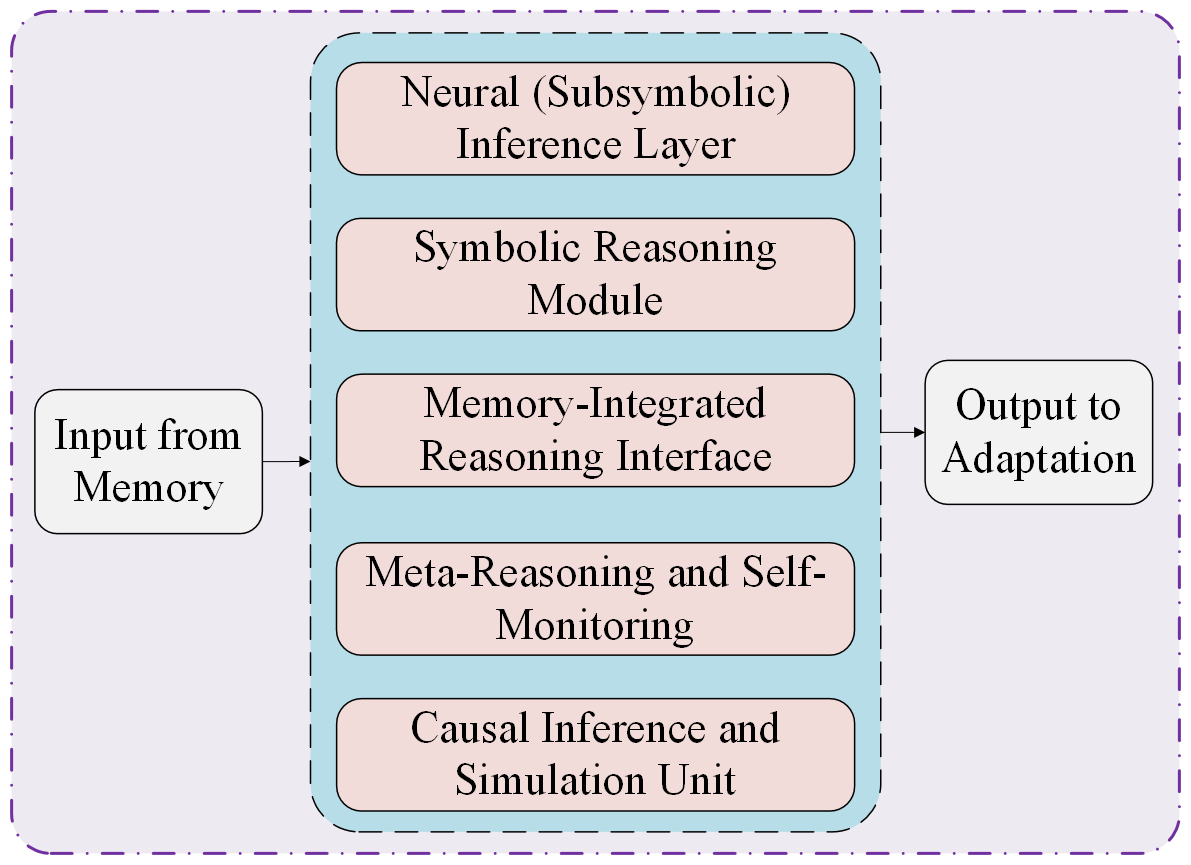}
  \caption{Reasoning and Inference Engine. A biologically grounded reasoning core that integrates neural inference layer, symbolic reasoning module, memory integrated reasoning inference, meta-reasoning, and causal inference.}
  \label{fig:reason}
\end{figure}

Architecturally, the module is a biologically inspired hybrid that combines subsymbolic neural inference \cite{besold2021neural} with symbolic rule-based reasoning \cite{garcez2008neural}, reflecting dual-process accounts of human cognition. At lower levels, sub-symbolic inference relies on embeddings, vector similarities, and latent projections for quick, associative predictions, which are effective for pattern recognition and handling noisy or incomplete inputs. These processes mirror distributed representations in the cortex and basal ganglia that support rapid, habit-like responses. At higher levels, symbolic mechanisms such as logic programs, ontologies, planning graphs, constraint solvers, and causal models facilitate structured reasoning, multi-step plan creation, contradiction detection, and counterfactual simulation, analogous to functions of the prefrontal and inferior frontal regions. The two layers operate together bidirectionally: sub-symbolic cues suggest possible hypotheses, while symbolic methods test, refine, and explain them.

A key feature of the engine is its close integration with Memory and the Learning Module. Episodic traces retrieved from Memory support case-based and analogical reasoning, while schemas and semantic relations offer abstract templates and causal priors. The Learning Module reveals these priors and heuristics as they are updated and, in turn, takes in artifacts from reasoning, such as inferred rules, counterfactuals, proofs, failures, and rationales, as additional training signals for consolidation and generalization. This process enables the system to transform deliberative insights into improved knowledge structures and decision policies.

The engine also supports meta-reasoning and self-reflection. It keeps an internal record of evidence usage, uncertainty, and assumption dependencies, enabling revisions when new data conflict with previous beliefs. These features are essential for reliability and accountability in safety-critical fields (e.g., clinical decision support or autonomous systems). In practice, these behaviors depend on uncertainty-aware inference, causal reasoning, counterfactual simulation, and neuro-symbolic components such as differentiable logic layers, knowledge-graph reasoning, and program-executing recurrent controllers that integrate learning and inference within the same computational loop. The engine is goal-driven and continually guided by the Attention Mechanism and Memory cues to determine how deep to reason, when to retrieve or simulate, and how to balance speed with deliberation.

In sum, the Reasoning and Inference Engine transforms the system from a passive recognizer into an active problem solver. As illustrated in Fig. \ref{fig:reason}, by integrating with the Memory–Learning complex along with attention and perception, it supports abstract thinking under uncertainty, analogical transfer, causal explanation, and plan creation within a closed-loop cognitive framework.

\subsection{Adaptation Layer}

The Adaptation Layer enables ongoing self-optimization and robustness. It modifies internal states, learning methods, and cognitive setups in response to environmental changes, new tasks, and shifting goals, based on principles such as neuroplasticity, neuromodulation, and executive reconfiguration (including anterior cingulate and prefrontal control). Unlike static pipelines, the architecture is designed for continuous operation: the Adaptation Layer updates parameters and routing without requiring full retraining, managing meta-learning, continual learning, and reinforcement learning, along with memory augmentation.

Functionally, the layer monitors prediction errors, outcome discrepancies, and uncertainty throughout the system, serving as a central controller for feedback. Importantly, it \emph{routes evaluative signals to the Learning Module and Memory}: reward, surprise, and error magnitudes influence representation updates, rehearsal scheduling, and consolidation; confidence and risk signals determine which reasoning strategies are used and how much deliberation is performed. This ensures the Memory–Learning complex stays aligned with task demands and the observed outcomes of actions.

Adaptation also involves strategic reconfiguration. It can reprioritize modalities and pathways (e.g., shift attention, trigger additional retrieval, or escalate from fast pattern matching to symbolic planning), or activate safety policies when uncertainty is high. When visual evidence becomes degraded, for example, adaptation might increase reliance on prior episodic traces or allocate more computation to simulation and counterfactual evaluation. The layer additionally maintains meta-representations of tasks and contexts to support transfer: it interfaces with Memory to fetch relevant experiences and with the Learning Module to adjust priors and skills for few-shot and zero-shot settings.

From an architectural perspective, the Adaptation Layer serves as a global modulator and scheduler that issues dynamic control signals to:
\begin{itemize}
    \item \textbf{Attention Mechanism}: re-weight salience and redirect focus;
    \item \textbf{Memory Module}: trigger consolidation, replay, and schema restructuring;
    \item \textbf{Learning Module}: set learning rates, select replay curricula, and apply plasticity constraints to avoid catastrophic forgetting;
    \item \textbf{Reasoning Engine}: choose inference style and depth (heuristic vs.\ deliberative, causal vs.\ associative);
    \item \textbf{Action Unit}: adapt decision policies, invoke defer/verify behaviors, or activate safety protocols under uncertainty.
\end{itemize}

\begin{figure}
  \centering
  \includegraphics[width=0.5\linewidth]{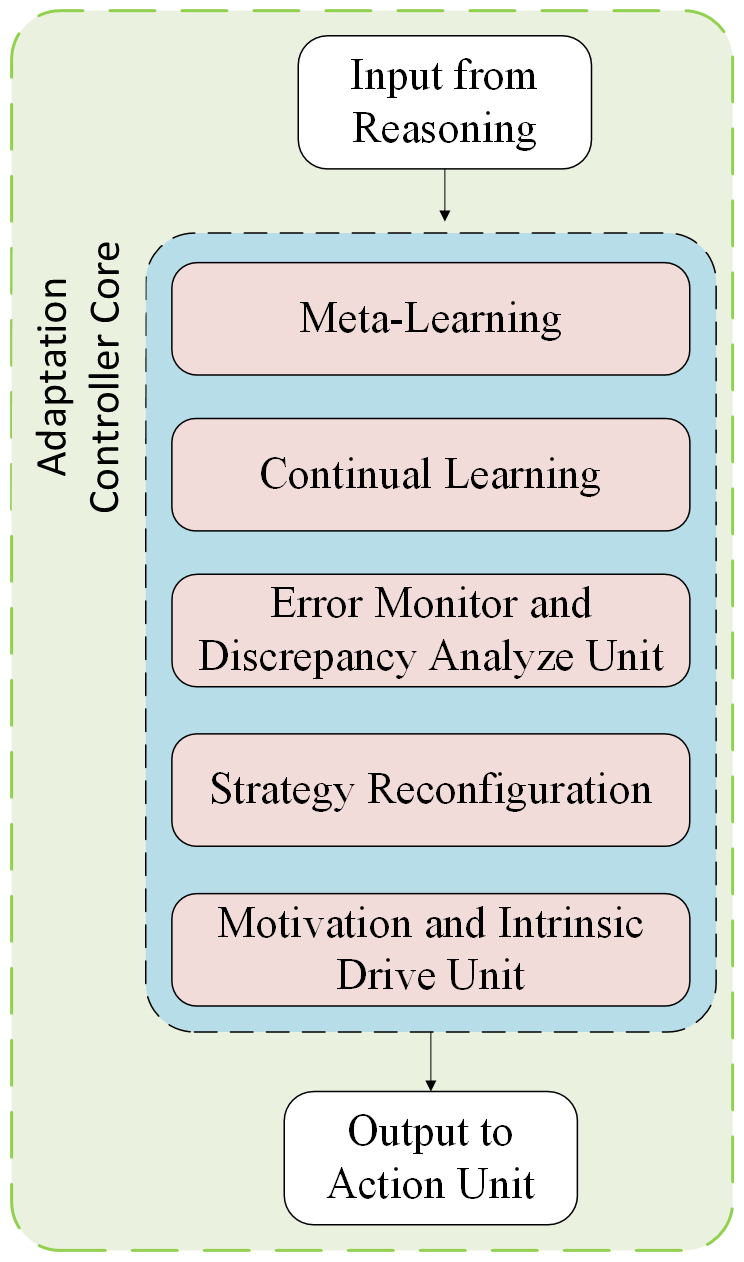}
  \caption{Adaptation layer. A meta-cognitive (thinking about cognition, i.e., monitoring, evaluating, and adjusting those processes) controller that enables dynamic reconfiguration, continual learning, and feedback-guided modulation across cognitive modules in response to prediction errors and environmental change.}
  \label{fig:adapt}
\end{figure}

In high-stakes or real-time applications, such as autonomous navigation, embodied robotics, or interactive medical diagnostics, the Adaptation Layer ensures long-term system durability by enabling safe exploration, rapid recovery from failures, and learning from sparse or delayed feedback. It also functions as the integrator for intrinsic motivation signals, such as novelty detection or curiosity-driven learning, allowing the system to explore and improve even without explicit supervision. Therefore, the Adaptation Layer transforms the neurocognitive architecture from a static model into a lifelong learning system. Continually aligning cognitive processes with task demands and environmental feedback helps the system stay competent, safe, and efficient over long periods and changing conditions. It forms the foundation of true general intelligence that learns how to learn, adapts its thinking, and evolves through experience. As shown in Fig. \ref{fig:adapt}, the adaptation layer acts as a meta-controller, modifying learning strategies, rerouting cognitive pathways, and monitoring uncertainty.

\subsection{Action and Decision Execution Unit}

The Action and Decision Execution Unit is the final point of cognitive processing in the neurocognitive intelligence framework. It converts the outputs of perception, attention, memory, and reasoning into concrete decisions, motor actions, or verbal responses based on the task context and system embodiment. More than a passive endpoint, this module actively influences learning, refines internal models, and supports goal-directed behavior through a continuous feedback loop.

At its core, this module is responsible for converting abstract representations into executable policies or actions. In an embodied system, like a robot, this may involve coordinated motor control, trajectory planning, or gesture execution. In contrast, for a non-embodied agent like a conversational assistant, the action might be generating contextually grounded language, structured outputs, or making predictive decisions. Importantly, the Action Unit is designed to remain sensitive to internal states such as uncertainty, intention, and context. Actions are not just static responses but are dynamically adjusted based on task complexity, environmental variability, and confidence signals from the reasoning module. One key feature of this module is its integration with goal representations and behavioral policies. Instead of executing a fixed response, the system uses goal-conditioned behavior \cite{ding2019goal}, which allows it modify the form, timing, or sequence of actions based on current objectives and long-term priorities. This is enabled by a feedback-aware architecture where actions are explicitly connected to goals maintained by the adaptation layer or executive controller. Through interaction with memory and attention systems, the agent can break down high-level goals into subtasks, choose suitable subroutines, and adapt its strategy if progress halts or unexpected results arise.

\begin{figure}
  \centering
  \includegraphics[width=0.85\linewidth]{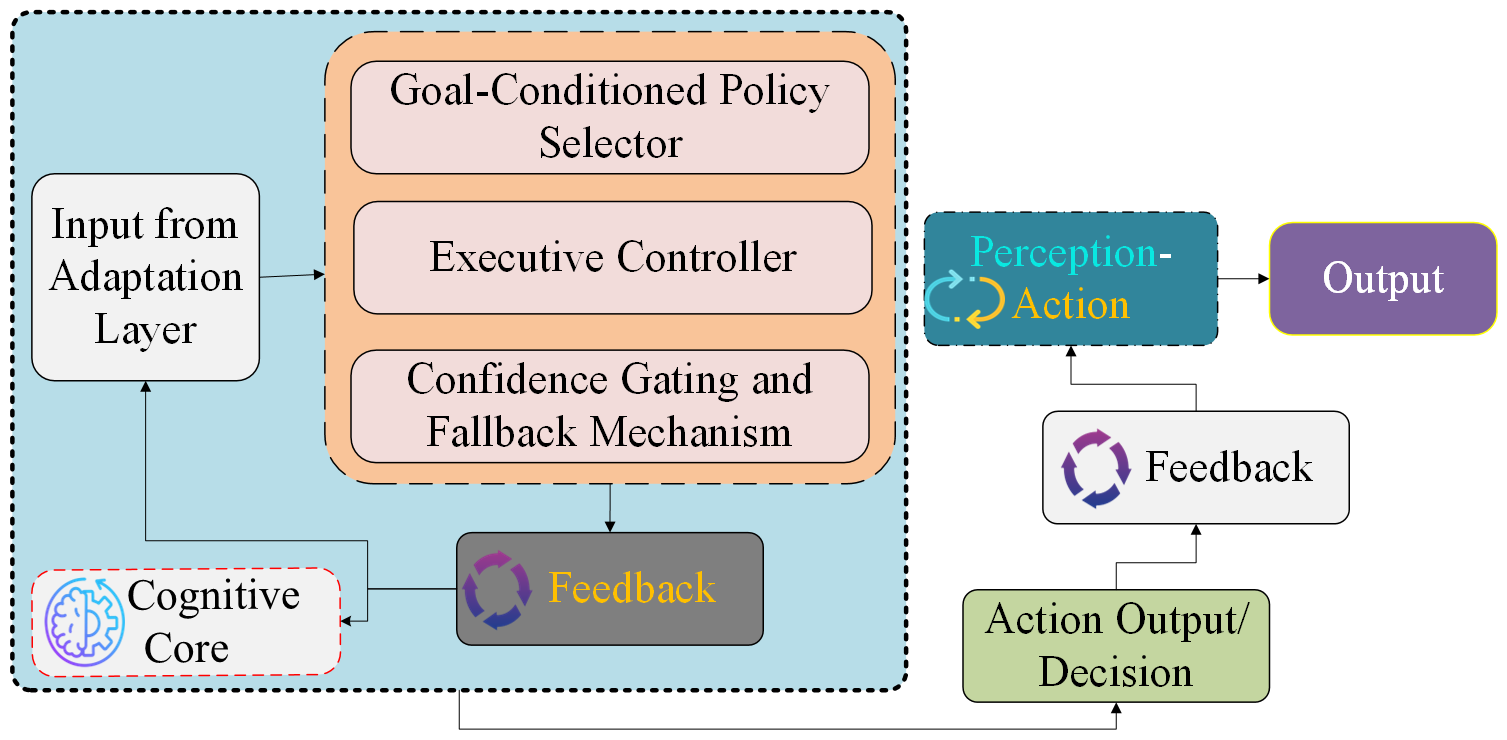}
  \caption{Action and Decision Execution Unit. A behavior-generation module that executes goal-conditioned actions (executing cognitive decisions as context-aware actions and transforming abstract intent into embodied responses.), integrates real-time feedback, and supports self-monitoring for continuous learning and robust interaction. By monitoring the effects of its outputs and comparing intended vs. actual outcomes, this module supports self-monitoring and feedback-driven adaptation, closing the perception–action loop.}
  \label{fig:act}
\end{figure}

Critically, action is not seen as a final step, but as a feedback mechanism that supports both internal and external adaptation. Each decision made produces internal effects, like success or failure signals and uncertainty levels, which are routed to the Adaptation Layer and Cognitive Core (e.g., attention, memory, reasoning) to refine internal strategies and control systems. At the same time, actions lead to environmental outcomes that are sensed externally and routed back through the Perception Unit, starting a new perception-action loop. These feedback pathways together support real-time learning, strategic reconfiguration, and ongoing adaptation. In biological systems, this principle is clear in how the brain continuously monitors motor results (via cerebellar loops) and evaluates outcomes (e.g., dopaminergic error signals). This structure mirrors those processes, turning each action into an opportunity for reflection and improvement. To promote self-improvement, the system tracks performance over time and records repeating success or failure patterns in episodic memory traces. This allows both reactive learning (immediate behavior adjustment based on negative feedback) and proactive planning (using stored knowledge to predict outcomes in new situations). The Action Unit thus acts as both an executor and a data source for overall system learning and adaptation.

Furthermore, the module includes mechanisms for confidence-based action gating and fallback options. When uncertainty is high or the predicted risk of error is unacceptable, the action layer can delay execution, request additional input, or revert to a safer, lower-risk policy. This is especially important in safety-critical domains such as healthcare robotics or autonomous driving, where blindly carrying out uncertain outputs could cause harm. Combining the Action Unit with meta-reasoning and adaptation signals ensures responsible decision-making in uncertain situations. In complex tasks, action sequences can be hierarchically organized, allowing for the execution of multistep behaviors with branching logic. This enables the agent to perform reactive decisions, procedural, and deliberative actions over extended periods. Integration with temporal reasoning, memory retrieval, and goal tracking ensures that these sequences remain coherent, adaptable, and grounded in real-world feedback. Therefore, the Action and Decision Execution Unit is a dynamic, feedback-aware controller that embodies the system’s decision-making capacity. Far more than just an output module, it plays a central role in grounding cognition in behavior, facilitating learning through feedback, and adjusting strategies toward flexible, goal-oriented intelligence. By linking reasoning, perception, attention, and memory with real-world execution, it closes the loop of neurocognitive intelligence, transforming thoughts into actions and actions into learning.

A key strength of the neurocognitive framework is its assumption of embodied cognition, the idea that intelligence is not purely abstract but rooted in interaction with the world. This aligns with current findings in cognitive science that highlight the importance of the body and environment in shaping mental processes \cite{wilson2002six}. When the framework adopts an embodied perspective, perception becomes deeply connected to the agent’s physical or simulated interactions with the environment. By incorporating sensorimotor feedback \cite{todorov2004optimality}, the framework gains access to real-time environmental constraints. This results in richer representations, context-aware decisions, and greater robustness to disturbances. The framework becomes situated, meaning it adapts to the context instead of relying solely on general rules or priors \cite{brooks1991intelligence}. Fig. \ref{fig:act} illustrates the final component of the system, the Action Unit, which converts internal cognition into real-world behavior and enables adaptive control through feedback.

To summarize the core modules of the neurocognitive architecture, their biological inspiration, functional roles, and AI analogs, we present the following mapping in Table \ref{table:nii_components}. This table shows how each component contributes to the looped system of cognitive control and intelligent behavior.

The previous sections have established the conceptual and architectural foundations of Neurocognitive-Inspired Intelligence (NII), emphasizing its biologically based modules for perception, memory, attention, reasoning, and adaptation. To place this contribution within the broader AI landscape, Table \ref{table:bii_xai_nii} offers a comparative overview of NII along with Brain-Inspired Intelligence (BII) and Explainable AI (XAI), illustrating how each approach addresses transparency, adaptability, and human-like cognition.

\newcolumntype{Y}{>{\RaggedRight\arraybackslash}X}

\begin{table*}[!t]
\caption{Mapping of Core Components in the Neurocognitive Intelligence Framework}
\label{table:nii_components}
\centering
\setlength{\tabcolsep}{6pt}
\renewcommand{\arraystretch}{1.15}
\begin{tabularx}{\textwidth}{|Y|Y|Y|Y|}
\hline
\rowcolor[HTML]{F2CEED}
\multicolumn{1}{|c|}{Component} &
\multicolumn{1}{c|}{Function in Human Cognition} &
\multicolumn{1}{c|}{Inspired by} &
\multicolumn{1}{c|}{AI Implementation} \\ \hline

Cognitive Control Loop &
Orchestrates perception, attention, memory, and action in a dynamic loop &
Prefrontal cortex, anterior cingulate cortex &
Central feedback loop coordinating modules \\ \hline

Perception \& Integration &
Sensory encoding, environmental modeling &
Sensory cortices (visual, auditory), posterior parietal cortex &
Multi-modal encoders (CNNs, transformers, symbolic fusion) \\ \hline

Attention &
Filters salient input, dynamically shifts focus based on goals or stimuli &
Dorsolateral prefrontal cortex, parietal lobe &
Transformer attention, saliency maps, dynamic routing \\ \hline

Memory &
Short- and long-term storage and retrieval of task-relevant data &
Working: prefrontal cortex; Episodic: hippocampus; Semantic: temporal and association cortices &
External memory modules, episodic stores, vector-based retrieval \\ \hline

Learning &
Extracts generalizable knowledge from experiences and updates stored representations and inference strategies &
Hippocampal–cortical plasticity, dopamine-driven reward prediction, offline replay &
Continual-learning algorithms, meta-learning with experience replay, structured representation updating \\ \hline

Reasoning \& Inference &
Logical thinking, abstraction, decision-making &
Prefrontal cortex, basal ganglia, inferior frontal gyrus &
Neuro-symbolic models, causal graphs, graph neural networks (GNNs) \\ \hline

Adaptation \& Modulation &
Monitors performance, modulates strategies, enables lifelong learning and recovery from failure &
Anterior cingulate cortex, neuromodulatory systems (dopaminergic, serotonergic) &
Meta-learning controllers, continual learning schedulers, strategic reconfiguration, feedback adaptation \\ \hline

Action \& Execution &
Goal-directed motor or verbal output &
Motor cortex, cerebellum, Broca’s area &
Decision policies, motor primitives, language decoders \\ \hline
\end{tabularx}
\end{table*}

\begin{table*}[!t]
\caption{Brain-Inspired Intelligence vs.\ XAI vs.\ Neurocognitive-Inspired Intelligence (NII)}
\label{table:bii_xai_nii}
\centering
\small  
\setlength{\tabcolsep}{5pt}
\renewcommand{\arraystretch}{1.15}
\begin{tabular}{|p{2.2cm}|p{4.2cm}|p{4.2cm}|p{4.2cm}|}
\hline
\rowcolor[HTML]{F2CEED}
\textbf{Aspect} & 
\textbf{Brain-Inspired Intelligence} & 
\textbf{Explainable AI (XAI)} & 
\textbf{Neurocognitive-Inspired Intelligence (NII)} \\ \hline
Definition &
AI models inspired by the structure of the brain (neurons, synapses, hierarchical layers). \emph{Looks like the brain} but doesn't think like it, often a black box. &
Techniques that explain the outputs of complex, often black-box AI models. \emph{Explains} what a black-box AI did after the fact. &
An AI framework inspired by the functional dynamics of human cognition, designed for built-in transparency, reasoning, and adaptation. \emph{Thinks and decides like the brain}, with built-in reasoning, explainability, and adaptability. \\ \hline
Focus &
Structural mimicry: recreating brain-like architectures (e.g., deep learning, spiking neurons). &
Interpretability: providing post-hoc explanations for otherwise opaque decisions. &
Functional cognition: emulating how humans perceive, reason, remember, and act. \\ \hline
Black-Box Status &
Often black-box (deep learning systems are inspired by brain structure but lack transparency). &
Starts from black-box models and adds explainability on top. &
Designed to be transparent from the ground up through explicit cognitive modules. \\ \hline
Goal &
Achieve intelligence by replicating biological architectures. &
Make AI outputs understandable to human users. &
Build human-like, adaptive, and interpretable intelligence that can reason, explain, and generalize. \\ \hline
Mechanism &
Deep neural networks, spiking neural models; structure-driven. &
Saliency maps, SHAP, LIME, decision trees; post-hoc or inherently interpretable models. &
Modular cognitive architecture with perception, memory, attention, reasoning, and adaptation, operating in closed-loop decision cycles. \\ \hline
Transparency &
Low: Internal processes are not easily interpretable. &
Medium to High: varies depending on the explainability method used. &
High: transparency is embedded in the decision-making process itself. \\ \hline
Adaptability &
Limited; often static after training. &
No direct adaptability; focused on explanation, not learning. &
Continuous learning, goal adaptation, real-time reasoning. \\ \hline
Example &
Deep learning models (CNNs, SNNs) trained on sensory data. &
Explaining why an image was classified as ``cat'' using feature importance. &
A robot or agent that explains its decisions while acting, reflects on outcomes, and adapts over time. \\ \hline
\end{tabular}
\end{table*}

As previously mentioned, Brain-Inspired Intelligence aims to mimic the brain's structural architecture, often leading to black-box models that lack interpretability and adaptability. Explainable AI (XAI) techniques attempt to mitigate this by offering post-hoc explanations or using interpretable model designs, but they usually operate on systems not originally built for transparency. In contrast, our proposed Neurocognitive-Inspired Intelligence (NII) framework is inherently transparent by design. By modeling distinct cognitive functions such as memory recall, attention modulation, and reasoning as explicit, modular components, NII offers explainable results, enabling agents to adapt, reason, and generalize in a human-like, context-aware manner. Therefore, while BII emphasizes structural mimicry and XAI focuses on explaining opaque decisions, NII focuses on making decisions that are explainable, goal-oriented, and cognitively grounded.

\begin{figure}
  \centering
  \includegraphics[width=0.85\linewidth]{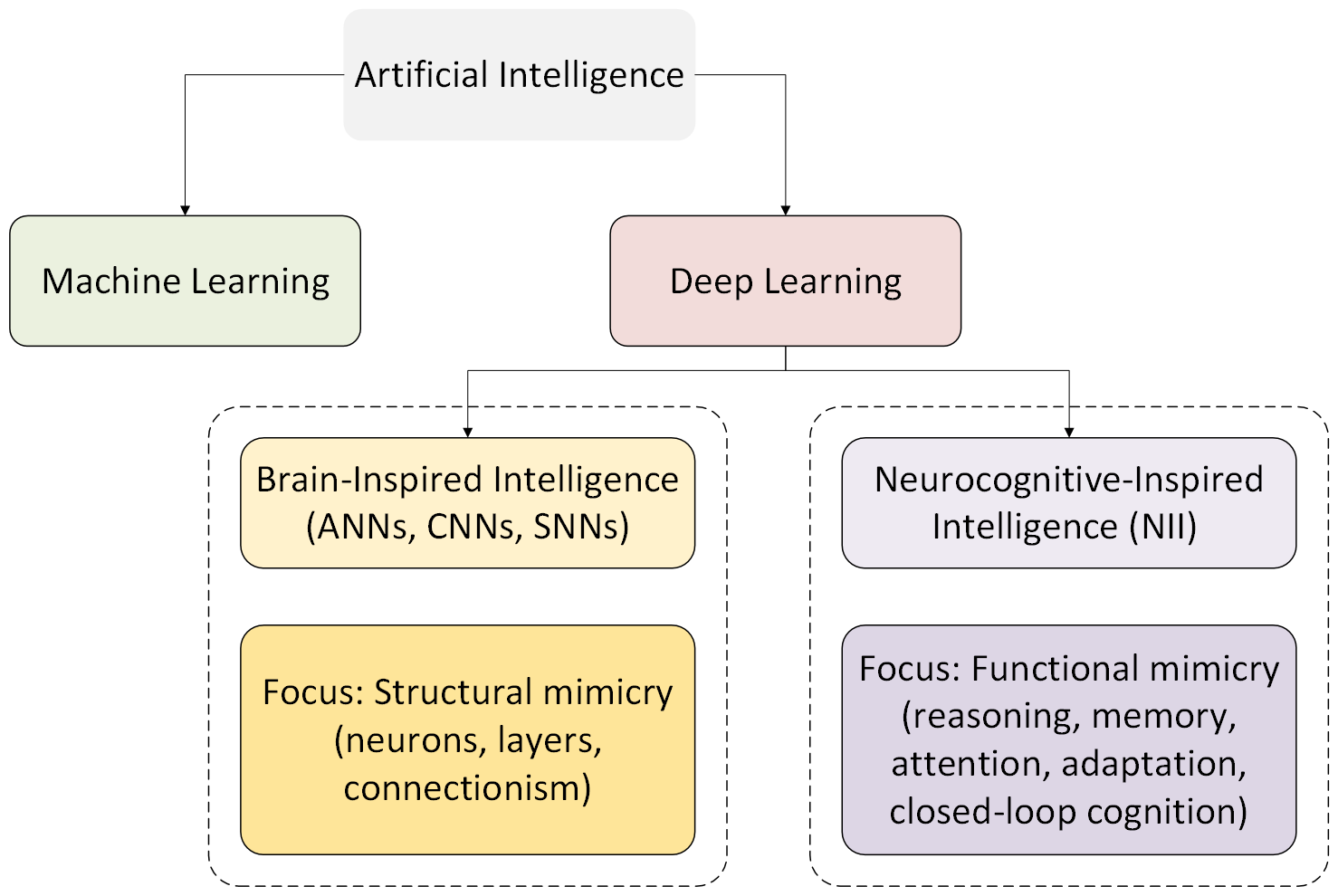}
  \caption{Positioning Neurocognitive-Inspired Intelligence (NII) within the AI Landscape. A conceptual taxonomy, illustrating the relationship between traditional machine learning, deep learning, brain-inspired intelligence, and the proposed NII paradigm. While Brain-Inspired Intelligence focuses on structural mimicry, NII builds upon and extends these approaches through functional mimicry of core cognitive faculties. This placement reflects a shift toward more explainable, human-aligned, and cognitively grounded AI.}
  \label{fig:fig_08}
\end{figure}

To place our proposed Neurocognitive-Inspired Intelligence (NII) within the broader AI landscape, Fig. \ref{fig:fig_08} presents a conceptual taxonomy of major AI paradigms, illustrating their relationships and fundamental principles. This hierarchy highlights how NII emerges as a distinct trajectory, building upon but exceeding the structural focus of Brain-Inspired Intelligence and the statistical basis of machine learning toward a more cognitively grounded and interpretable framework. The taxonomy is introduced at this point, after thoroughly explaining NII’s architectural components, to ensure its theoretical position and unique features are properly contextualized and understood.

Testing such frameworks in virtual cognitive environments, such as simulated 3D worlds or interactive text-based games, can validate their robustness, generalization, and planning abilities. These environments serve as cognitive sandboxes for verifying theory-grounded components under controlled complexity \cite{ha2018recurrent}. The neurocognitive framework outlined here synthesizes decades of cognitive science and neuroscience into an actionable architecture for next-generation AI. By integrating core functions such as perception, memory, attention, reasoning, and action into a unified loop, the model offers a promising path toward generalization, adaptability, and explainability, qualities that are notably absent in conventional AI.


\begin{figure*}[h]
  \centering
  \includegraphics[width=0.85\linewidth]{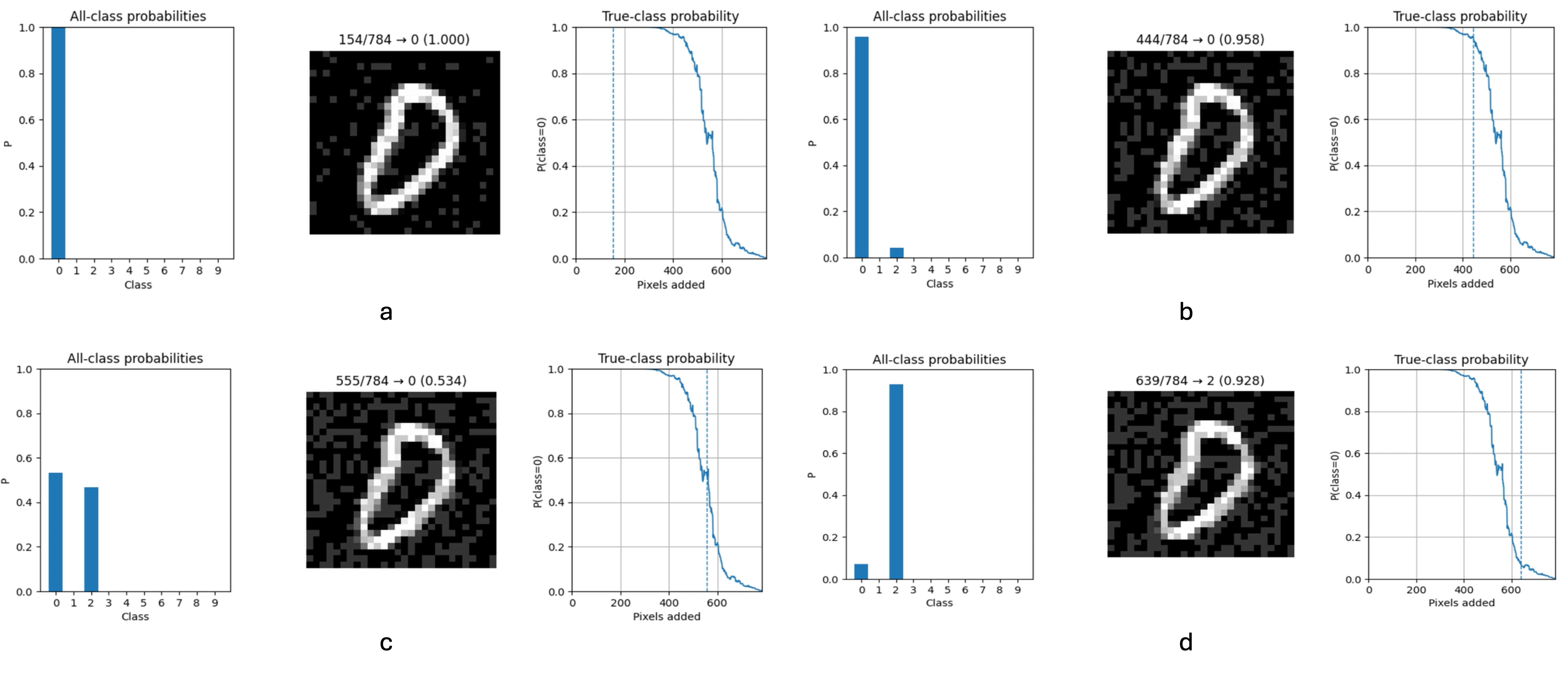}
  \caption{Monitoring adversarial perturbation effects and demonstrating how the number of pixels added affects classification confidence in a CNN: As more pixels are added, classification probability shifts progressively toward the adversarial target class a: 19\% of pixels are added to the clean image. b: 56\% of pixels are added to the clean image. c: 70\% of pixels are added to the clean image. d: 81\% of pixels are added to the clean image, demonstrating cumulative adversarial effects and increasing the probability of targeted class.}
  \label{fig:mnist}
\end{figure*}


\begin{figure*}[h]
  \centering
  \includegraphics[width=0.52\linewidth]{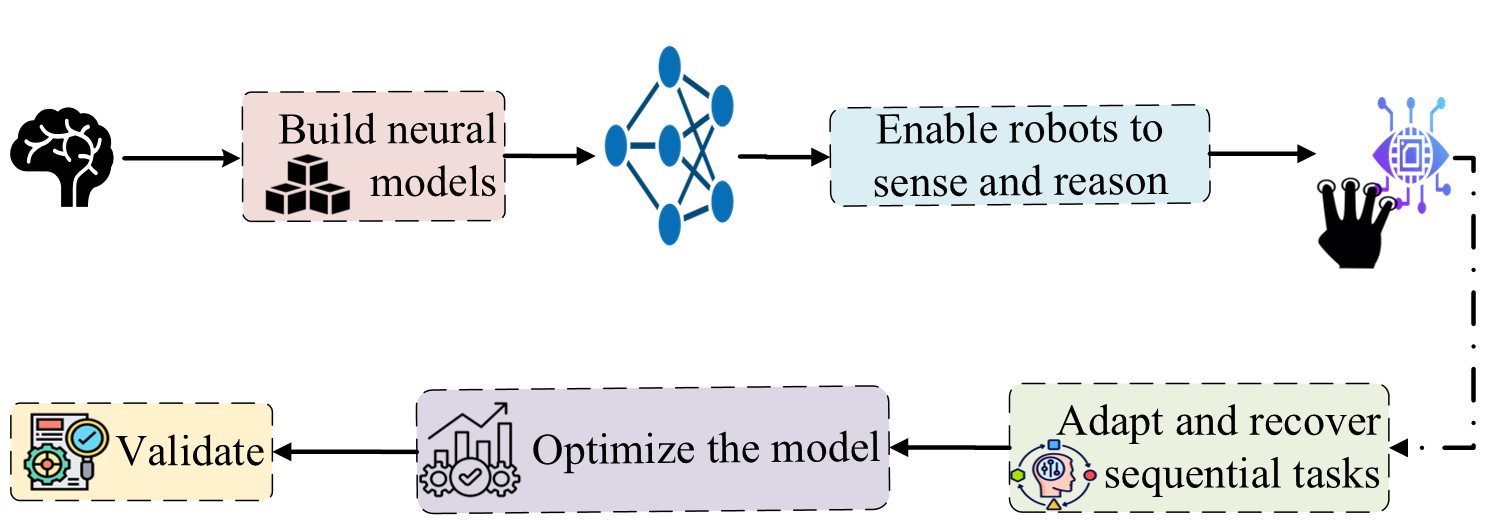}
  \caption{Multitask robot intelligence pipeline based on neurocognitive AI. The process begins by building neural models inspired by brain-like processing. These models are trained to enable robots to sense and reason about their environment using multimodal sensory input. Once equipped with core perception and cognition capabilities, robots are designed to adapt and recover during the execution of complex, sequential tasks, including manipulation or navigation. Optimization steps refine the neural models based on performance feedback, while validation ensures the system’s generalizability, safety, and robustness. This cyclical pipeline allows the robotic agent to self-improve over time, handle task variability, and recover from failure, all while maintaining efficient and scalable computation.}
  \label{fig:multi_task}
\end{figure*}



\section{From Framework to Function: Implementation, Case Studies, and Scalability Challenges}
\label{discussion}

This section details pathways for implementing the proposed neurocognitive framework, including use cases across major domains, evaluation methods, and scalability considerations. This comprehensive overview shows how theory turns into practice, highlighting both opportunities and challenges in achieving neurocognitive intelligence.

\subsection{System Integration and Modular Design}

A robust neurocognitive architecture requires a modular structure, allowing each functional unit can be independently optimized and seamlessly integrated. Biological systems, especially the human brain, depend on this distributed processing model where cortical and subcortical regions performing specialized yet interconnected roles. For example, the SPAUN model \cite{stewart2012spaun}, built using the Nengo simulator \cite{bekolay2014nengo}, demonstrates how high-level cognition can develop from basic spiking neural activity through modular interactions \cite{eliasmith2012large}. Similarly, the LEABRA (Local, Error-driven and Associative, Biologically Realistic Algorithm) model \cite{o1996leabra} combines Hebbian and error-driven learning to mimic cognitive behaviors by integrating cortical and subcortical functions \cite{o2000computational}. Neuromorphic hardware, such as Intel’s Loihi chip \cite{davies2018loihi}, enables the deployment of such models in energy-efficient settings by simulating neural spikes for real-time, event-driven computation \cite{davies2018loihi}. On the software front, frameworks such as ROS-Neuro \cite{tonin2022ros} facilitate real-time, biologically inspired communication protocols, enabling plug-and-play connectivity between neural and cognitive modules across distributed systems.

\subsection{Learning Strategies}

\subsubsection{Online and Continual Learning}

A major challenge in AI is catastrophic forgetting, which is the loss of previously acquired knowledge when learning new tasks. Neurocognitive intelligence tackles this by using dual-memory systems, where the hippocampus stores short-term memory and the neocortex handles long-term consolidation. This biologically inspired method is applied in AI through techniques like Elastic Weight Consolidation (EWC), which penalizes updates to parameters essential for previous tasks \cite{kirkpatrick2017overcoming}. Another approach, Synaptic Intelligence, tracks the importance of weights during learning and gradually reduces the plasticity of key parameters over time \cite{zenke2017continual}. These continual learning methods mimic human-like adaptability and help prevent disruptive overwriting of neural weights when learning new data streams.

\subsubsection{Few-Shot and Meta-Learning}

Humans excel at learning with minimal data, relying on prior experience and generalizable knowledge. Few-shot learning in neurocognitive systems can be achieved using memory-augmented neural networks (MANNs), which enable internal storage and retrieval of contextual examples \cite{santoro2016meta}. Meta-learning, or “learning to learn,” allows models to optimize their learning process based on past tasks, reducing the data needed for future performance \cite{finn2017model}. These approaches mirror the brain’s ability to generalize quickly by transferring knowledge across domains, which is essential for real-world applications such as clinical decision support or adaptive robotics.

\subsection{Evaluation Metrics for Neurocognitive Intelligence}

Conventional metrics such as accuracy or F1-score may fall short when assessing systems designed to emulate the cognitive flexibility and generalization capabilities of human intelligence. Instead, neurocognitive evaluation should emphasize dimensions that reflect the system’s ability to adapt, reason, and behave in cognitively aligned ways across dynamic contexts. Neurocognitive evaluation must incorporate generalization, memory retention, attention dynamics, explainability, and embodied interaction. These criteria provide a more comprehensive view of performance in both simulated and real-world settings. Testbeds like the Animal-AI Olympics \cite{crosby2020animal}, which assess agents in environments designed to mimic animal intelligence tests, and BabyAI \cite{chevalier2019babyai}, a grid-world platform for instruction-following under uncertainty, offer cognitively challenging benchmarks. Additionally, biologically grounded evaluation can use EEG or Event Related Potential (ERP) alignment to verify that AI decisions mimic human neural responses \cite{ren2024brain}. These metrics ensure that AI systems are not only functionally correct but also cognitively plausible.

\subsection{Proof-of-Concept and Applications}

To demonstrate the practical significance and applicability of the proposed conceptual framework, we present targeted proof-of-concept experiments that demonstrate key functional principles of neurocognitive-inspired intelligence, specifically, adaptive perception under perturbation and cross-modal integration through tactile-visual learning.

\subsubsection{Tracking Pixel-Level Perturbations and Dynamic Classification Adaptation: A Step Toward Context-Sensitive Learning and Neurocognitive-Inspired Intelligence}

To examine the vulnerability of convolutional neural networks (CNNs) \cite{lecun2002gradient} to adversarial perturbations, we conducted an experiment using the Fast Gradient Sign Method (FGSM) noise \cite{goodfellow2014explaining}, incrementally modifying pixel values while observing changes in the classification probabilities of the true and target classes (see Fig. \ref{fig:mnist}) \cite{deng2012mnist}. Unlike conventional studies that assess overall model degradation, our approach tracks each added pixel dynamically, analyzing its direct impact on classification confidence. In this experiment, FGSM noise was generated and applied incrementally, with one randomly selected pixel added to the image at each step. The pixels were chosen arbitrarily from the noise distribution and integrated into the image without a predefined pattern. As more pixels were introduced, the probability of the true class progressively decreased, while the probability assigned to an adversarially targeted class increased. Crucially, this experiment demonstrates a real-time monitoring mechanism akin to adaptive reasoning, where each pixel alteration shapes subsequent predictions. This suggests an early-stage capability for context-sensitive learning, reinforcing the potential for neurocognitive-inspired AI systems that adjust their decision-making dynamically in response to changing stimuli.

\subsubsection{Resilient Multistep Robotic Intelligence}
Inspired by the structural and functional organization of the brain, we examine the design of resilient robotic systems capable of learning, adapting, and recovering from failure during complex task execution (see Fig. \ref{fig:multi_task}). The approach can incorporate cortical-like architectures to support adaptive recall, error correction, and long-range spatiotemporal reasoning. By combining vision and touch through crossmodal attention, these robots can perform tactile shape completion and affordance-aware manipulation even under occlusion or uncertainty. While recent studies have successfully integrated visual and tactile modalities for tasks such as surface \cite{kansana2025surformerv1, kansana2025surformerv2} and object classification or grasp refinement
\cite{zhang2023visual} \cite{ding2023adaptive} \cite{corradi2017object}, these efforts typically focus on narrowly defined goals within a limited pipeline. In contrast, the proposed framework goes beyond sensory fusion to model the broader loop of cognitive processing, all inspired by neurobiological principles, positioning our work not merely as a multimodal system but as a step toward resilient neurocognitive robotics capable of general-purpose, multitask adaptation in dynamic environments.

This framework may extend beyond recognition or control by addressing the entire cognitive loop, including mechanisms for hierarchical learning, anomaly detection, and feedback-driven adaptation. It enables real-time diagnosis and replanning during multistep operations. To facilitate practical deployment, neuromorphic approximations and model compression techniques are used for energy-efficient and hardware-neutral execution. Overall, this neurocognitive framework shows how brain-inspired architectures and multimodal integration can produce autonomous agents capable of performing complex sequential tasks and remaining resilient against unpredictable challenges.

To illustrate the importance of neurocognitive principles in multisensory robotic perception, we conducted an experiment examining the interaction between vision and touch in material classification. Using the touch and go dataset \cite{yang2022touch}, we implemented three CNN-based classification pipelines: vision-only, tactile-only, and a combined multimodal configuration. The batch size was 16 for the tactile modality and 64 for both the vision and combined modalities. A consistent learning rate of 0.0001 was used across all configurations over 100 epochs. Early stopping with a patience of 15 epochs based on validation accuracy was employed. For the vision and tactile modalities, the learning rate was adaptively reduced using the ReduceLROnPlateau scheduler, with a factor of 0.8 and a patience of 5 epochs. For the combined modality, a separate early stopping criterion with a reduced patience of 10 epochs was applied to balance training stability and performance.

The dataset contains 4,909 samples in total, which were split into 80\% for training and 20\% for testing. We used a filtered subset of the full GelSight dataset, selecting fewer but diverse classes with the following class distribution: 
Concrete (383), Plastic (74), Wood (552), Metal (105), Brick (695), Tile (317), Synthetic Fabric (406), Tree (169), Grass (749), Rock (972), and Plants (487). 

\begin{figure*}[h]
  \centering
  \includegraphics[width=0.6\linewidth]{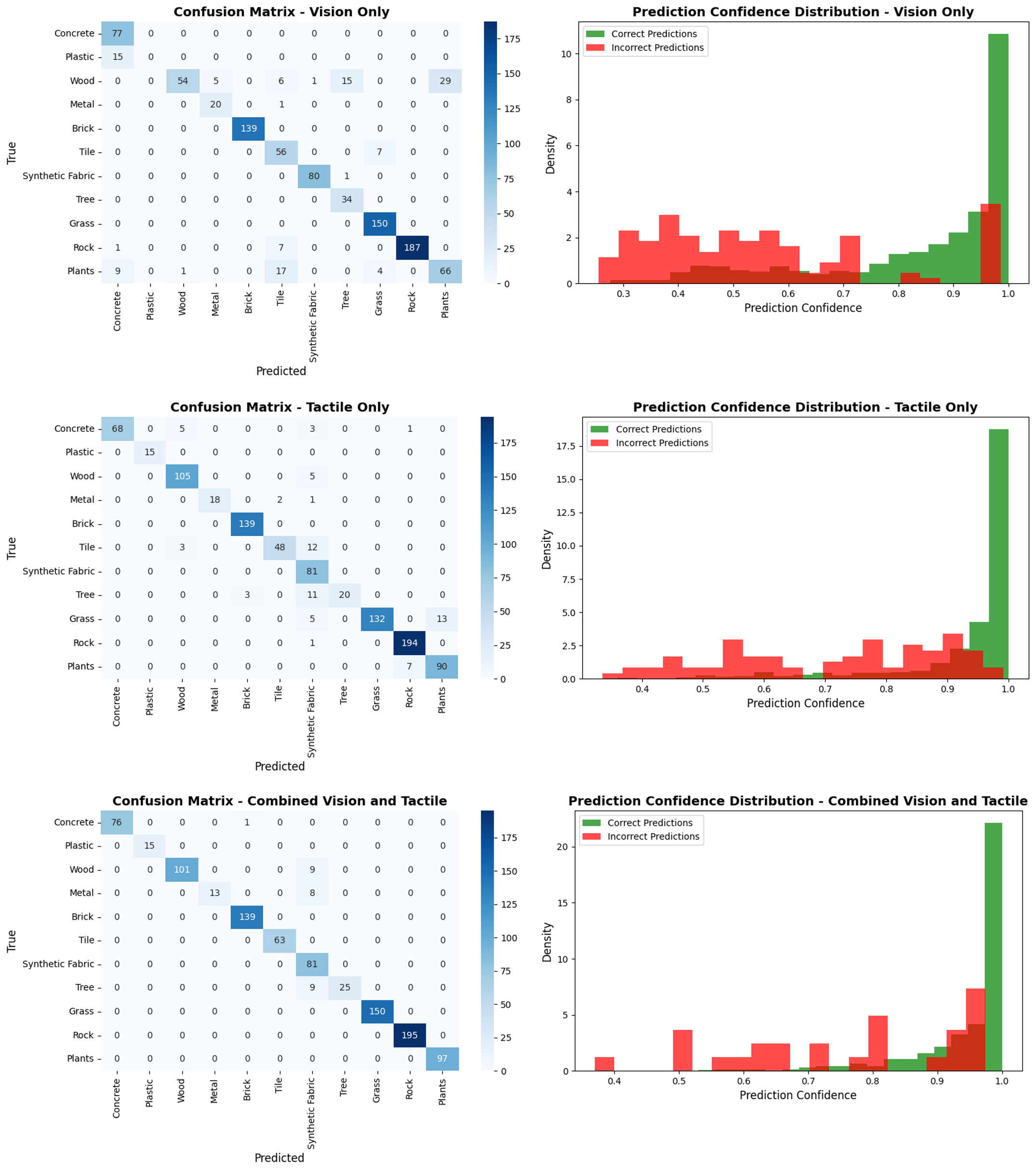}
  \caption{Confusion matrices and prediction confidence distributions for CNN-based surface classification using three input conditions: Vision-only input (top), Tactile-only input (middle), Combined vision and tactile input (bottom). The left column shows confusion matrices for each setting, while the right column displays the corresponding prediction confidence distributions. The combined model exhibits the highest classification accuracy and confidence, indicating that multimodal integration improves robustness and discriminative power.}
  \label{fig:cnn_eval}
\end{figure*}

Moreover, as shown in Fig. \ref{fig:cnn_eval}, the confusion matrix for the \textbf{vision-only model} indicates moderate performance with an overall accuracy of \textbf{87\%}, although it has difficulty distinguishing certain surface classes like Tree and Concrete. In comparison, the \textbf{tactile-only model} achieves \textbf{92\% accuracy} (see Table \ref{table:cnn_eval}), with better distinction across most categories. This suggests that when tactile information is gathered effectively, it can outperform vision-only models, especially in scenarios with occlusions or unclear visual cues, emphasizing the important role of touch in supporting reliable perception in challenging conditions.

\begin{table}[!t]
\caption{Evaluation results of CNN model}
\label{table:cnn_eval}
\centering
\setlength{\tabcolsep}{8pt}
\renewcommand{\arraystretch}{1.15}
\begin{tabular}{|l|c|c|c|c|}
\hline
\rowcolor[HTML]{F2CEED}
\textbf{Modality} & \textbf{Accuracy} & \textbf{F1} & \textbf{Precision} & \textbf{Recall} \\ \hline
Vision Only  & 0.8788 & 0.7781 & 0.7704 & 0.8145 \\ \hline
Tactile Only & 0.9267 & 0.9006 & 0.9489 & 0.8852 \\ \hline
Combined     & 0.9726 & 0.9474 & 0.9773 & 0.9327 \\ \hline
\end{tabular}
\end{table}

Most notably, the \textbf{combined vision and tactile model} surpasses both unimodal systems, achieving an accuracy of \textbf{97\%} and showing higher confidence along with fewer misclassifications across all categories. This supports the idea that multimodal fusion enhances generalization and robustness, which are key goals in neurocognitive system design.

\begin{figure*}[h]
  \centering
  \includegraphics[width=0.7\linewidth]{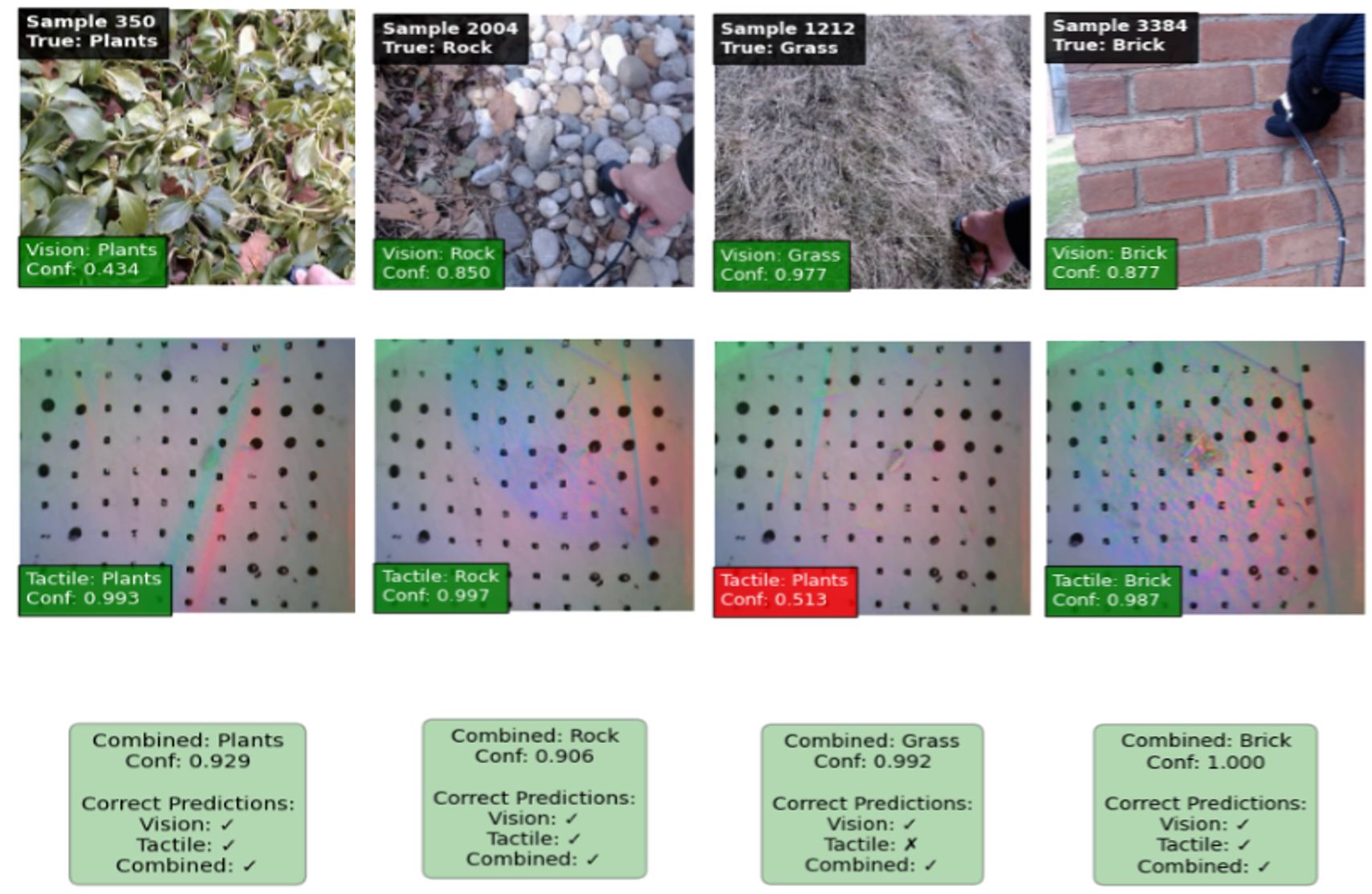}
  \caption{Multimodal surface classification using concatenated visual and tactile inputs. The top row displays predictions from the vision-only model, which correctly classifies Plants, Rock, Grass, and Brick with high confidence. The middle row shows results from the tactile-only model. The bottom row presents results from the combined model (vision + tactile), which successfully corrects the misclassification and consistently achieves accurate predictions across all samples, demonstrating the benefit of multimodal integration.}
  \label{fig:multi_modal}
\end{figure*}

To further explore the combined strengths of vision and tactile senses in surface classification, we performed a multimodal experiment by concatenating GelSight tactile images with visual data, achieving an overall accuracy of 97\%. This method demonstrates a proof of concept for sensory integration, showing how even simple concatenation can improve performance and robustness in line with the neurocognitive framework. Fig. \ref{fig:multi_modal} displays the surface classification results from our multimodal test, where vision and tactile (GelSight) data were combined to boost predictive accuracy. The top row shows visual inputs for four samples (Rock, Grass, Plants, Brick) with their respective predictions and confidence levels from the vision-only model. For example, Sample 350 (True: Plants) is correctly identified as Plants with a confidence of 0.434, and Sample 2004 (True: Rock) is predicted as Rock with a confidence of 0.850. The middle row illustrates the tactile (GelSight) images for these samples, capturing surface textures through contact-based imaging. The tactile-only model accurately predicts Brick (Sample 3384, Conf: 0.987) and Rock (Sample 2004, Conf: 0.997), but misclassifies Grass as Plants (Sample 1212, Conf: 0.513), highlighted in red, likely due to surface similarity or tactile pattern similarities. The bottom row shows the combined (vision + tactile) predictions, achieving high accuracy: Plants (Conf: 0.929), Rock (Conf: 0.906), and Brick (Conf: 1.000) are correctly identified, and the Grass sample (Conf: 0.992) is also correctly predicted, overcoming the tactile model’s misclassification.

While this experiment demonstrates that integrating tactile and visual information enhances surface classification accuracy, its primary contribution lies in illustrating the relevance of multisensory perception to the broader neurocognitive robotics paradigm. Unlike this feedforward architecture, the neurocognitive framework would incorporate biologically inspired mechanisms for memory, reasoning, and adaptive control, enabling real-time decision-making and lifelong learning in complex, uncertain environments. Such an approach lays the groundwork for developing resilient, multistep robotic intelligence.
Creating resilient multistep robotic intelligence based on this framework has the potential to function effectively even in situations of ambiguity or incomplete information. For example, by physically interacting with and touching part of an object, a robot can infer its structure or identity using learned crossmodal representations, thereby extending perception beyond what is immediately visible. This adaptable ability enables neurocognitive systems to generalize across different environments and recover from partial sensory loss, making this experiment an early yet significant step toward more robust, embodied, and context-aware intelligence.

Additionally, the neurocognitive framework can be customized for various real-world fields, including healthcare, manufacturing, education, and robotics. Although these applications vary in context, they all rely on the same core architecture consisting of perception, attention, memory, reasoning, adaptation, and action modules. Importantly, the following items represent initial conceptual designs that establish the foundation for future implementations and research efforts. They show how NII can improve decision-making, resilience, and adaptability in safety-critical and dynamic environments.

\subsubsection{Cognitive Health Monitoring in Aging Populations}

This approach can be used in eldercare settings for early detection of dementia and mild cognitive impairment. By analyzing daily behavioral patterns and speech variations, AI agents can identify anomalies that suggest cognitive decline \cite{konig2015automatic}. For instance, sensor data combined with neuro-symbolic reasoning \cite{bhuyan2024common} can be used to detect deviations from routine behaviors in real time. A diagram showing how the framework components relate to this use case is included in Table \ref{table:aging_monitoring_mapping}, illustrating how each layer helps in early identification of cognitive decline.

\newcolumntype{Y}{>{\RaggedRight\arraybackslash}X}

\begin{table*}[!t]
\caption{Mapping Framework Components to Cognitive Health Monitoring in Aging Populations}
\label{table:aging_monitoring_mapping}
\centering
\setlength{\tabcolsep}{8pt}
\renewcommand{\arraystretch}{1.15}
\begin{tabular}{|p{3.5cm}|p{12cm}|}
\hline
\rowcolor[HTML]{F2CEED}
\textbf{Framework Component} & \textbf{Application in Cognitive Health Monitoring} \\ \hline
Perception Unit & Captures multimodal data such as motion sensor logs, audio recordings, and video feeds. \\ \hline
Attention Mechanism & Prioritizes significant behavioral or speech changes, e.g., pauses, hesitations, or disorientation. \\ \hline
Memory Module & Stores historical behavior and speech data for temporal analysis and long-term monitoring. \\ \hline
Reasoning Engine & Applies rules and pattern recognition to infer deviations from cognitive norms. \\ \hline
Adaptation Layer & Adjusts thresholds for anomaly detection based on individual routines and health status. \\ \hline
Action/Output Unit & Triggers alerts/reports to caregivers when early warning signs of cognitive decline are detected. \\ \hline
\end{tabular}
\end{table*}

\subsubsection{Industrial Safety and Hazard Prediction}

\begin{figure*}[h]
  \centering
  \includegraphics[width=0.6\linewidth]{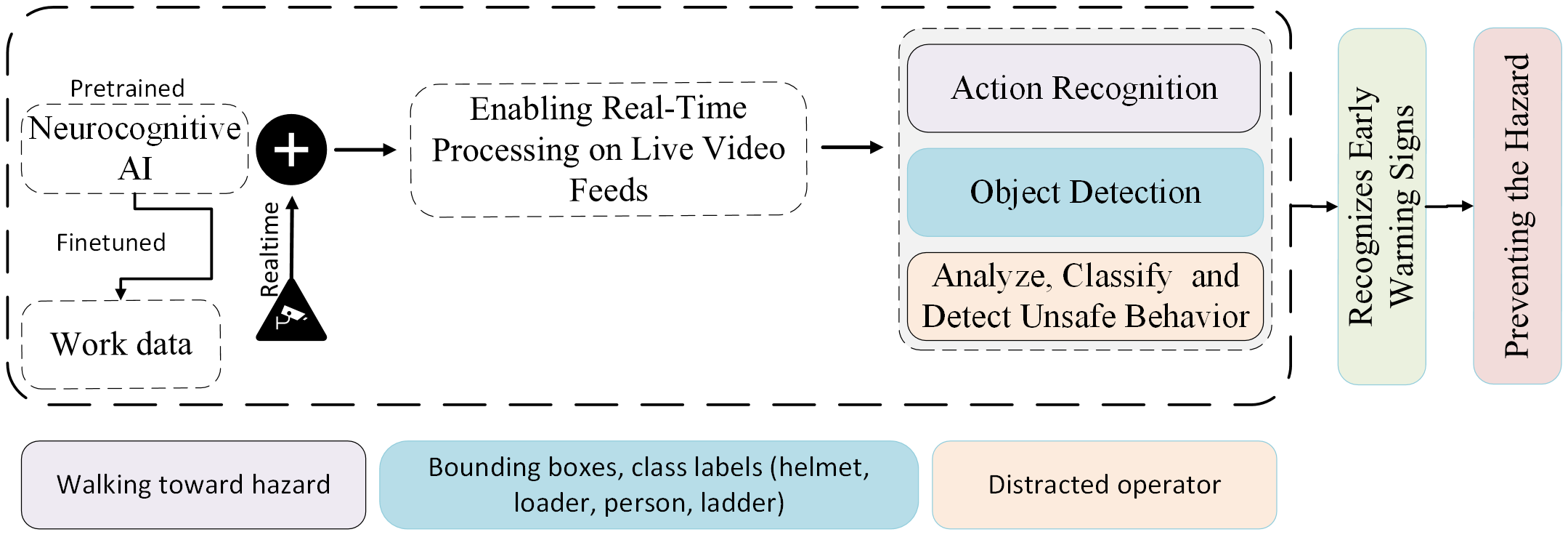}
  \caption{Real-time deployment pipeline of neurocognitive AI for safety-critical smart manufacturing environments.
The system begins with a pretrained neurocognitive AI model that is fine-tuned using task-specific work data. This refined model is deployed for real-time processing on live video feeds captured from the work environment. The processing pipeline consists of three core components: (1) Action Recognition to understand worker behavior, (2) Object Detection to localize and track relevant tools or machinery, and (3) Unsafe Behavior Analysis to classify and detect hazardous interactions or patterns. These components together recognize early warning signs of potential hazards. By continuously adapting to dynamic environments, the system enables proactive safety interventions to prevent accidents before they occur.
}
  \label{fig:deploy}
\end{figure*}

In smart manufacturing environments, neurocognitive AI can predict unsafe human behavior or equipment failure by analyzing real-time, multi-modal sensor data from equipment, environment, and human operators (see Fig. \ref{fig:deploy}). Systems with predictive attention modules and hierarchical planning frameworks have demonstrated improved safety compliance \cite{webster2011environmental} \cite{bobadilla2014predictive} \cite{ozobu2025advancing}. Unlike traditional automation, these agents adjust their responses, attention, and reasoning to continuously changing conditions based on shifting environmental cues, allowing for proactive risk detection and mitigation of risks \cite{golcarenarenji2022machine} \cite{ahn2023safefac}. 

As mentioned earlier, the key to this capability is the integration of \textbf{attention mechanisms} that focus on the most critical safety signals, such as unusual machine vibrations, operator fatigue cues, or unexpected environmental changes. This allows the system to allocate computational resources efficiently and escalate urgent alerts. The \textbf{memory module} keeps track of contextual histories of equipment performance and worker behavior, supporting trend analysis and anomaly detection over time. Paired with a \textbf{reasoning engine} capable of causal inference and predictive modeling, these systems can predict potential failures or unsafe behaviors before incidents happen. Additionally, the \textbf{adaptation layer} continuously adjusts safety thresholds and intervention policies based on environmental feedback and incident outcomes, ensuring the system remains responsive to both overall safety standards and site-specific risks. The \textbf{action unit} converts these insights into real-time alerts, automated safety protocols, or direct commands to robotic systems, effectively completing the loop on hazard prevention.

By incorporating neurocognitive principles, such as embodied cognition through sensorimotor integration and top-down attention modulation, these AI systems develop a more nuanced understanding of complex, dynamic industrial environments, promoting safer and more resilient manufacturing ecosystems.

\begin{table*}[!t]
\caption{Key Scalability and Deployment Challenges for Neurocognitive Systems}
\label{table:nii_scalability}
\centering
\setlength{\tabcolsep}{6pt}
\renewcommand{\arraystretch}{1.15}
\begin{tabularx}{\textwidth}{|Y|Y|Y|}
\hline
\rowcolor[HTML]{F2CEED}
\multicolumn{1}{|c|}{Challenge} &
\multicolumn{1}{c|}{Description} &
\multicolumn{1}{c|}{Emerging Solutions} \\ \hline

Hardware limitations &
High computational and memory demands &
Neuromorphic hardware (e.g., Loihi, SpiNNaker) \cite{enuganti2025neuromorphic,davies2021advancing}; edge AI optimization \cite{surianarayanan2023survey} \\ \hline

Lack of multimodal datasets &
Sparse datasets that combine vision, language, memory, and motor data &
Benchmarks like BabyAI \cite{chevalier2018babyai,chevalier2019babyai}, DynaBench \cite{kiela2021dynabench} \\ \hline

Integration complexity &
Difficulty combining symbolic, neural, and sensorimotor modules &
Neuro-symbolic compilers \cite{zhang2025neuro} \\ \hline

Simulation constraints &
Real-time simulation of perception--action loops is hard to scale &
Frameworks like NengoDL \cite{rasmussen2019nengodl}, OpenCog AtomSpace \cite{belachew2018shifting} \\ \hline

Continual learning challenges &
Catastrophic forgetting in lifelong scenarios &
Elastic weight consolidation \cite{liu2020incdet}; memory-augmented networks \cite{santoro2016meta}; rehearsal strategies \cite{shaughnessy1981memory} \\ \hline
\end{tabularx}
\end{table*}

\subsubsection{Personalized Learning in Education}

Cognitive tutors that adapt based on user memory retention, attention span, and conceptual understanding can transform e-learning. Neurocognitive systems can model student mental states, including attention dynamics, working memory capacity, and conceptual mastery, and adjust teaching methods accordingly. Unlike traditional adaptive learning systems that mainly depend on static performance metrics, neurocognitive tutors dynamically analyze multi-modal learner data, such as interaction patterns, response times, and physiological signals (e.g., eye tracking, heart rate), to develop a comprehensive understanding of student engagement and knowledge retention. A key component of this approach is the \textbf{attention mechanism}, which detects when a learner’s focus diminishes or when specific concepts need reinforcement, allowing the system to adjust task difficulty or deliver timely prompts. The \textbf{memory module} tracks learner progress over time, supporting spaced repetition and retrieval practices to improve long-term retention. Meanwhile, the \textbf{reasoning engine} uses neuro-symbolic models to identify misconceptions, provide personalized feedback, and create individualized learning paths. Recent efforts using neuro-symbolic architectures in platforms like ASSISTments \cite{heffernan2014assistments} demonstrate improved knowledge retention and engagement \cite{akyuz2020effects} \cite{sung2023analysis} \cite{zhang2017evaluating} \cite{koedinger2006cognitive}. 

The \textbf{adaptation layer} further personalizes the experience by including contextual factors such as prior knowledge, emotional state, and environmental distractions, which affect learning effectiveness. Lastly, the \textbf{action/output unit} provides tailored exercises, real-time hints, or encouragement, creating an interactive dialogue with the learner that mimics human tutoring interactions. By leveraging the neurocognitive framework’s focus on dynamic feedback loops, embodied cognition, and context-aware reasoning, AI-powered personalized learning systems can deliver scalable, high-quality education customized to each student's unique needs and abilities.

\subsection{Scalability and Deployment Challenges}

Scalability remains a major challenge in deploying neurocognitive systems in real-world environments. These systems encounter architectural, computational, and data-related issues that must be resolved for widespread adoption. Table \ref{table:nii_scalability} outlines key challenges, from hardware limitations to integration and continual learning, and features emerging solutions from current research and toolkits. Toolkits like OpenCog \cite{hart2008opencog} and NengoDL \cite{rasmussen2019nengodl} are working to close this gap by providing scalable infrastructure for building cognitive architectures with real-time simulation capabilities. As shown, deploying neurocognitive intelligence practically requires robust architectures, continual and few-shot learning methods, advanced evaluation techniques, and interdisciplinary collaboration.

\section{Conclusion}
\label{conclusion}

As artificial intelligence continues to evolve, it is increasingly evident that current architectures are fundamentally constrained in their ability to demonstrate general intelligence, adapt across different tasks, reason over time, and operate effectively in real-world environments. These constraints, from data inefficiency to poor generalization and absence of lifelong learning, highlight a significant gap between engineered AI systems and the natural intelligence exhibited by biological systems. This paper argues that bridging this gap requires a shift toward neurocognitive-inspired intelligence, a paradigm inspired not only by biological neurons but also by the full range of cognitive faculties, which include memory, perception, attention, reasoning, adaptation, and action that support natural intelligence. Drawing on principles from neuroscience, cognitive science, and systems biology, we propose a biologically inspired conceptual framework that organizes these cognitive functions into a cohesive, dynamic loop. Our hybrid framework emphasizes three core architectural features: (1) the integration of symbolic and sub-symbolic reasoning to enable both statistical pattern recognition and logical abstraction, (2) the embodiment of cognition through sensorimotor feedback loops to promote grounded and adaptive learning, and (3) the alignment of architectural design with principles derived from neuroscience to support modularity, hierarchy, and biological plausibility. We also discuss enabling technologies such as neuromorphic computing and cognitive simulation platforms, which are essential for scaling these architectures in practice. However, this work does not present a complete system but rather a theoretical blueprint, a biologically inspired pathway toward more general and resilient forms of artificial intelligence. The architecture we propose is intended as a research agenda, inviting future empirical exploration, implementation, and validation. Achieving the full potential of neurocognitive intelligence will require ongoing interdisciplinary collaboration, technological advancement, and a shared commitment to grounding AI in the mechanisms of natural cognition. It urges us to reconsider not only how we build intelligent machines but also how we define intelligence itself. Just as the invention of neural networks reshaped decades of symbolic AI, the neurocognitive intelligence framework could define the next era of artificial intelligence, where machines not only compete but also observe, plan, evaluate, and even experience in human-like ways. The journey ahead will be challenging and nonlinear. Yet, by grounding our architectures in the principles of natural cognition and fostering integrative, biologically grounded research, we move closer to building machines that do not merely imitate intelligence but begin to sense and understand it.



\end{document}